\documentclass[12pt]{article}
\usepackage{amssymb,makeidx}
\usepackage[dvips]{graphicx}
\usepackage{cite}
\newcommand{\be}{\begin{eqnarray}}
\newcommand{\ee}{\end{eqnarray}}
\newcommand{\bez}{\begin{eqnarray*}}
\newcommand{\eez}{\end{eqnarray*}}
\newcommand{\pa}{\partial}
\newcommand{\na}{\nabla}

\renewcommand{\d}{{\rm d}}
\newcommand{\dl}{\delta}
\newcommand{\ad}{{\rm ad}}
\newcommand{\D}{{\rm D}}
\newcommand{\A}{{\cal A}}
\renewcommand{\L}{\pounds}
\newcommand{\X}{{\cal X}}

\renewcommand{\O}{\Omega}
\newcommand{\oA}{\otimes_\A}
\newcommand{\U}{{\cal U}}
\newcommand{\V}{{\cal V}}
\newcommand{\R}{{\cal R}}
\newcommand{\lc}{\lrcorner\,}
\newcommand{\E}{\mathfrak{E}}

\title{\bf Differential Geometry of Group Lattices}
\date{  }
\author{{\bf Aristophanes Dimakis}\thanks{Electronic mail: dimakis@aegean.gr}  \\
 Department of Financial and Management Engineering, \\
 University of the Aegean, 31 Fostini Str., GR-82100 Chios
 \and
 {\bf Folkert M\"uller-Hoissen}\thanks{Electronic mail: fmuelle@gwdg.de}\\
 Max-Planck-Institut f\"ur Str\"omungsforschung, \\
 Bunsenstrasse 10, D-37073 G\"ottingen
}

\begin{document}
\maketitle

\renewcommand{\theequation} {\arabic{section}.\arabic{equation}}
\newtheorem{theorem}{Theorem}[section]
\newtheorem{lemma}{Lemma}[section]

\newcounter{example}[section]
\renewcommand{\theexample}{\arabic{section}.\arabic{example}}
\newenvironment{example}
{ \refstepcounter{example} \noindent
  {\it Example \arabic{section}.\arabic{example}}.}
{ \hfill $\blacksquare$ \vspace{.2cm} }

\vskip1cm

\begin{abstract}
In a series of publications we developed ``differential geometry'' on
discrete sets based on concepts of noncommutative geometry.
In particular, it turned out that first order differential calculi
(over the algebra of functions) on a discrete set are in bijective
correspondence with digraph structures where the vertices are given
by the elements of the set. A particular class of digraphs are Cayley
graphs, also known as group lattices. They are determined by a discrete
group $G$ and a finite subset $S$. There is a distinguished
subclass of ``bicovariant'' Cayley graphs with the property
$\ad(S)S \subset S$.

We explore the properties of differential calculi which arise
from Cayley graphs via the above correspondence.
The first order calculi extend to higher orders and then
allow to introduce further differential geometric structures.

Furthermore, we explore the properties of ``discrete'' vector fields
which describe deterministic flows on group lattices. A Lie
derivative with respect to a discrete vector field and an inner
product with forms is defined. The Lie-Cartan identity then holds
on all forms for a certain subclass of discrete vector fields.

We develop elements of gauge theory and construct an analogue
of the lattice gauge theory (Yang-Mills) action on an arbitrary
group lattice. Also linear connections are considered and a
simple geometric interpretation of the torsion is established.

By taking a quotient with respect to some subgroup of the discrete
group, generalized differential calculi associated with so-called
Schreier diagrams are obtained.
\end{abstract}

\newpage

\tableofcontents

\newpage

\section{Introduction}
\setcounter{equation}{0}
In a series of papers
\cite{DMHS93,DMH94_fs,DMH94_gr,DMHV95,DMH96,BDMHS96,DMH99-dRg} we
developed differential geometry on discrete sets (see also Refs.
\cite{Conn+Lott90,Sita95,Cast,Majid,Dai+Song01,Mack01,Mall+Rapt01}
for related work).
A key concept is a differential calculus (over the algebra $\A$ of functions)
on a set. First order differential calculi on discrete sets were found to
be in bijective correspondence with digraph structures \cite{DMH94_gr},
where the vertices of the digraph are given by the elements of the set and
neither multiple arrows nor loops are admitted.
In particular, this supplies the elements of the set with neighborhood
relations. An important example is a differential calculus which
corresponds to the hypercubic lattice and which leads to an elegant
formulation of lattice gauge theory \cite{DMHS93}.
\vskip.1cm

A special class of digraphs are \emph{Cayley graphs}\cite{Cayley}
(see Refs. \cite{Bollo98,MKS76}, for example), which are also known as
\emph{group lattices} in the physics literature. These are determined
by a discrete group $G$ and a subset $S$. The elements of $G$ are the
vertices of the digraph and the elements of $S$ determine (via right
action) arrows from a vertex $g$ to ``neighboring'' vertices.
Hypercubic lattices, on which the usual lattice (gauge) theories are
built, are special Cayley graphs. Another example of importance
for physics is the truncated icosahedron which models the $C_{60}$
Fullerene \cite{Samu92}. Physical models on group lattices have also
been considered in Refs. \cite{Samu91,Lech+Samu95,Lech+Samu96,Regge+Zecc96},
in particular.
Furthermore, Cayley graphs play a role in the study of connectivity
and routing problems in communication networks (see Ref. \cite{Heyd+Duco97}
for a review).
\vskip.1cm

The above-mentioned correspondence between digraphs and first order
differential calculi suggests to explore those calculi
which correspond to Cayley graphs.
Moreover, given a first order differential calculus which corresponds
to a Cayley graph, it naturally extends to higher orders so that
we have a notion of $r$-forms, $r > 1$. This provides the basis for
introducing further differential geometric structures, following
general recipes of noncommutative geometry.
\vskip.1cm

In section~\ref{sec:fodc} we introduce first order differential
calculi associated with group lattices. Our approach very much
parallels standard constructions in ordinary differential geometry.
In particular, we first introduce vector fields on a group lattice
and then 1-forms as duals of these.
Section~\ref{sec:diffmaps} concerns maps between group lattices
which are ``differentiable'' in an algebraic sense \cite{DMHV95}.
Of special importance for us are ``bicovariant'' group lattices
$(G,S)$ with the property that the left and right actions on
$G$ with respect to all elements of $S$ is differentiable.
\vskip.1cm

A first order differential calculus naturally extends to higher orders,
i.e. to a full differential calculus. The structure of differential
calculi obtained from group lattices is the subject of
section~\ref{sec:hodc}.
\vskip.1cm

Geometric relations are often more conveniently expressed in terms
of vector fields than forms. In section~\ref{sec:discr_vf} we introduce
a special class of vector fields which we call ``discrete'' and a
subclass of ``basic'' vector fields and explore their properties.
A Lie derivative with respect to a discrete vector field and
an inner product of discrete vector fields and forms is defined.
For basic vector fields with differentiable flow the Lie-Cartan
formula holds.
\vskip.1cm

Section~\ref{sec:connections} treats connections on (left or right)
$\A$-modules over differential calculi associated with group lattices.
In particular, Yang-Mills fields are considered and
an analogue of the lattice gauge theory action on an arbitrary
group lattice is constructed.
\vskip.1cm

If the module is the space of 1-forms, we are dealing with linear
connections. This is the subject of section~\ref{sec:lincon}.
In particular, we find that the condition of vanishing torsion
of a linear connection has a simple geometric meaning.
\vskip.1cm

A differential calculus on a group lattice induces a ``generalized
differential calculus'' on a coset space. The resulting differential
calculus is generalized in the sense that the space of 1-forms
is, in general, larger than the $\A$-bimodule generated by the image
of the space of functions under the action of the exterior derivative.
There is a generalized digraph (``Schreier diagram'' \cite{Bollo98})
associated with such a first order differential calculus which
in general has multiple links and also loops.
Some further remarks are collected in section~\ref{sec:concl}.

\section{First order differential calculus associated with a
         group lattice}
\label{sec:fodc}
\setcounter{equation}{0}
Let $G$ be a discrete group and $\A$ the algebra of complex-valued
functions $f : G \to \mathbb{C}$.\cite{genform}
With $g \in G$ we associate $e^g \in \A$ such that
$e^g(g') = \dl_{g,g'}$ for all $g' \in G$. The set of
$e^g$, $g \in G$, forms a linear basis of $\A$ over $\mathbb{C}$,
since every function $f$ can be written in the form
$f = \sum_{g \in G} f(g) \, e^g$. In particular, we have
$e^g \, e^{g'} = \dl^{g,g'} \, e^g$ and
$\sum_{g \in G} e^g = \mathbf{1}$, where $\mathbf{1}$ denotes
the constant function which is the unit of $\A$.
\vskip.1cm

The left and right translations by a group element $g$,
$L_g(g') = g g'$ and $R_g(g') = g' g$, induce
automorphisms of $\A$ via the pull-backs
$(L^\ast_g f)(g') = f(L_g g') = f(gg')$
and $(R^\ast_g f)(g') = f(R_g g') = f(g'g)$.
In particular, we obtain
\be
   L^\ast_g  e^{g'} = e^{g^{-1}g'} \, , \qquad \qquad
   R^\ast_g  e^{g'} = e^{g' g^{-1}}   \label{R^ast-e^g}
\ee
for all $g,g' \in G$. Introducing\cite{notation}
\be
  \ell_g f = R^\ast_g f - f
\ee
so that $(\ell_g f)(g')=f(g'g)-f(g')$,
we find the modified Leibniz rule
\be
  \ell_g (f f') = (\ell_g f)(R^\ast_g f') + f (\ell_g f') \, .
\ee
The maps $\ell_g : \A \to \A$, $g \in G$, generate an
$\A$-bimodule via
\be
  (f \cdot \ell_g) f' := f \, \ell_g f' \, , \qquad
  (\ell_g \cdot f) f' := (\ell_g f')(R^\ast_g f)
\ee
so that
\be
  \ell_g \cdot f = (R^\ast_g f) \cdot \ell_g \, .
\ee
Indeed, one easily verifies that
\be
  (ff') \cdot \ell_g = f \cdot (f' \cdot \ell_g) \, , \qquad
  \ell_g \cdot (ff') = (\ell_g \cdot f) \cdot f' \, .
\ee
The modified Leibniz rule can now be written as
\be
   \ell_g (f f') = (\ell_g \cdot f') f + (f \cdot \ell_g) f' \; .
\ee
\vskip.1cm

Let $S$ be a {\em finite} subset of $G$ which does not contain the
unit of $G$. From $G$ and $S$ we construct a directed graph as follows.
The vertices of the digraph represent the elements of $G$ and
there is an arrow from the site (vertex) representing $g$ to the one
representing $gh$ if and only if $h \in S$. In other words,
there is an arrow from $g$ to $g'$ iff $g^{-1} g' \in S$.
A digraph obtained in this way is called a {\em Cayley graph} or a
{\em group lattice}.\cite{infG}

\begin{lemma}
\label{lemma:ccu}
The connected component of the unit $e$ in the
group lattice is the subgroup of $G$ generated by $S$.
\end{lemma}
{\bf Proof:} Let $H$ be the subgroup of $G$ generated by $S$.
Every element $g \in H$ can be written as a finite product
$g = h_1^{k_1} \cdots h_r^{k_r}$ with $h_i \in S$ and
$k_i \in \{ \pm 1 \}$. If $k_r = 1$, there is an arrow
from $h_1^{k_1} \cdots h_{r-1}^{k_{r-1}}$ to $g$.
If $k_r = - 1$, there is an arrow
from $g$ to $h_1^{k_1} \cdots h_{r-1}^{k_{r-1}}$. By
iteration, $g$ is connected to $e$.
Hence $H$ is contained in the connected component
${\cal C}_e$ of $e$. Because of the group property,
every element connected to an element of $H$ must
itself be an element of $H$. Hence ${\cal C}_e = H$.
\hfill $\blacksquare$
\vskip.1cm

It follows that the group lattice $(G,S)$ is connected if
and only if $S$ generates $G$ (see also Ref. \cite{Serr80}, p.17).
If the subgroup $H$ generated by $S$ is smaller than $G$,
the group lattice consists of a set of disjoint but isomorphic
parts corresponding to the set of left cosets $g H$, $g \in G$.
\vskip.1cm

 For $h \in S$, the maps $\ell_h : \A \to \A$ are naturally
associated with the arrows of the digraph since
$(\ell_h f)(g) = f(gh) - f(g)$
is the difference of the values of a function $f$ at two
connected ``neighboring'' points of the digraph.
The maps $\ell_h$ generate an $\A$-bimodule $\X$.\cite{subbimodule}
At each $g \in G$, they span a linear space which we call
the {\em tangent space} at $g$.
\vskip.1cm

Let $\O^1$ be the $\A$-bimodule dual to $\X$ such that
\be
   \langle f \cdot X, \alpha \rangle
 = \langle  X, f \alpha \rangle
 = f \langle X, \alpha \rangle \, , \qquad
   \langle X \cdot f, \alpha \rangle
 = \langle X, \alpha f \rangle    \label{bracket_prop}
\ee
for all $X \in \X$, $f \in \A$ and $\alpha \in \O^1$.
If $\{ \theta^h | h \in S\}$ denotes the set of elements of $\O^1$
dual to $\{ \ell_h | h \in S \}$, so that
$\langle \ell_{h'}, \theta^h \rangle = \dl^h_{h'}$, then
\be
   \langle \ell_{h'}, \theta^h f \rangle
 = \langle \ell_{h'} \cdot f, \theta^h \rangle
 = \langle (R^\ast_{h'} f) \cdot \ell_{h'}, \theta^h \rangle
 = R^\ast_{h'} f \, \dl^h_{h'}
 = \langle \ell_{h'}, (R^\ast_h f) \theta^h \rangle
\ee
for all $h,h' \in S$. Hence
\be
   \theta^h \, f = R^\ast_h f \, \theta^h \, .
          \label{theta_f}
\ee
The space of 1-forms $\O^1$ is a free $\A$-bimodule and
$\{\theta^h | \, h \in S \}$ is a basis.
A linear map $\d : \A \to \O^1$ can now be introduced by
\be
  \d f = \sum_{h \in S} (\ell_h f) \, \theta^h \, .
         \label{df_ell_h}
\ee
It satisfies the Leibniz rule $\d(ff') = (\d f) f' + f (\d f')$.
In particular, we obtain
\be
   \d e^g = \sum_{h \in S}(\ell_h e^g) \, \theta^h
 = \sum_{h \in S}(e^{gh^{-1}}-e^g) \, \theta^h  \; .
         \label{de_theta}
\ee
Now we multiply both sides from the left by $e^{gh^{-1}}$ with
some fixed $h \in S$. Since $h$ is different from the unit element
of $G$, we obtain $e^{gh^{-1}} \d e^g = e^{gh^{-1}} \theta^h$.
 From this we find\cite{G-infinite}
\be
  \theta^h = \sum_{g \in G} e^{gh^{-1}} \d e^g
           = \sum_{g \in G} e^g \, \d e^{gh} \, .
\ee
Furthermore,
\be
  \theta := \sum_{h \in S} \theta^h
          = \sum_{g \in G, \, h \in S} e^g \, \d e^{gh}
            \label{theta_def}
\ee
satisfies
\be
  \d f = \theta f - f \theta = [\theta,f] \, .
             \label{df_theta}
\ee
Moreover, we obtain
\be
    \langle X, \d f \rangle = X f \; .  \label{X_df}
\ee
\vskip.1cm

Let us introduce
\be
  {\cal I} = \{ (g,g') \in G \times G \, | \, g^{-1} g'
                \not \in S_e \}
\ee
where $S_e = S \cup \{ e \}$.
This is the set of pairs $(g,g')$ for which $e^g \, \d e^{g'} = 0$.
Note that $e^g \, \d e^g = - e^g \, \theta \neq 0$.
\vskip.1cm

The first order differential calculus $(\A,\O^1,\d)$ constructed
above is also obtained from the universal first order
differential calculus $(\A,\O^1_u,\d_u)$ as the quotient
$\O^1 = \O^1_u/{\cal J}^1$ with respect to the submodule
${\cal J}^1$ of $\O^1_u$ generated by all elements of the
form $e^g \, \d_u \, e^{g'}$ with $(g,g') \in {\cal I}$.
If $\pi_u : \O^1_u \to \O^1$ denotes the corresponding
projection, then we have $\d = \pi_u \, \d_u$.

\begin{lemma}
\label{lemma:subgroup-univdc}
If $S_e$ is a subgroup of $G$, the corresponding
first order differential calculus on the component connected to
the unit is the universal one.
\end{lemma}
{\bf Proof:} According to Lemma~\ref{lemma:ccu}, the $e$-component is $S_e$.
Since for every pair $(h,h') \in S_e \times S_e$, $h \neq h'$,
there is an element $h'' \in S$ such that $h = h' h''$,
there is an arrow from $h'$ to $h$ in the associated digraph.
Hence all pairs of different elements of $S_e$ are connected
by a pair of antiparallel arrows. This characterizes the
universal differential calculus.
\hfill $\blacksquare$
\vskip.2cm

\begin{example}
\label{ex:G=Z,S=1}
One of the simplest examples is obtained as
follows. Let $G = \mathbb{Z}$, the additive group of
integers, and $S = \{1 \}$. Then we have
$(\ell_1 f)(k) = f(k+1)-f(k)$ and
$\theta^1 = \sum_{k \in \mathbb{Z}} e^k \, \d e^{k+1}$.
Introducing the coordinate function
$t = \sum_{k \in \mathbb{Z}} k \, e^k$, we find
$\theta^1 = \d t$ and $\ell_1 f = \pa_{+t} f$ with the
discrete derivative $\pa_{+t}f(t) = f(t+1)-f(t)$.
Hence
\be
      \d f = (\pa_{+t}f) \, \d t \; .
\ee
This example is important as a model for a discrete parameter
space, and in particular as a model for discrete time.
A generalization is obtained by taking the additive group
$G = \mathbb{Z}^n$ and
$S = \{ (1,0,\ldots,0), (0,1,0,\ldots,0), \ldots, (0,\ldots,0,1) \}
   =: \{ \hat{m} \, | \, 1 \leq m \leq n \}$
which generates $G$.
This leads to an oriented hypercubic lattice digraph. Then
$(\ell_{\hat{m}} f)(k) = f(k+\hat{m})-f(k) =: (\pa_{+\hat{m}} f)(k)$
and
$\theta^{\hat{m}} = \sum_{k \in \mathbb{Z}^n} e^k \, \d e^{k+\hat{m}}$.
Introducing coordinates via
$x = \sum_{k \in \mathbb{Z}^n} k \, e^k = (x^1, \ldots, x^n)$,
we find
\be
   \d f = \sum_{m=1}^n (\pa_{+\hat{m}} f) \, \d x^m \, ,
   \qquad
   \theta^{\hat{m}} = \d x^m  \; .
\ee
This differential calculus appeared first in Ref. \cite{DMHS93}
(see also Ref. \cite{DMH92}) and turned out to be useful, in particular,
in the context of lattice gauge theory \cite{dc+lgt} and completely
integrable lattice models \cite{DMH96}.
\end{example}

\begin{example}
\label{ex:G=Zm,S=1}
Let $G = \mathbb{Z}_m$ ($m=2,3, \ldots$), the finite additive group
of elements $0,1,2,m-1$ with composition law addition modulo $m$.
The unit element is $e=0$. Choosing $S = \{ 1 \}$, we have a single
basis 1-form $\theta^1$. In contrast to example~\ref{ex:G=Z,S=1},
here $\theta^1$ is not exact. Indeed, suppose that $\theta^1 = \d f$
for some function $f$. This is equivalent to $\ell_1 f = \mathbf{1}$
which leads to the contradiction $m = \sum_{g} (\ell_1 f)(g) = 0$.
By taking direct products of this lattice, a group lattice structure
for $G = \mathbb{Z}_m^n$ is obtained.
\end{example}

\begin{example}
\label{ex:G=Z2,Z3,Z4}
For $G = \mathbb{Z}_2$, the only group lattice is the complete digraph
corresponding to the universal first order differential calculus on
the two elements $\{ 0,1 \}$. For $G = \mathbb{Z}_3$, one has to
distinguish two cases. If $S$ contains a single element only,
the group lattice is a closed linear chain of arrows
(cf example~\ref{ex:G=Zm,S=1}).
The choice $S = \{ 1,2 \}$ leads to the complete digraph on the three
elements and thus to the universal differential calculus. Less simple
structures appear for $G = \mathbb{Z}_m$, $m > 3$.
For example, choosing $G = \mathbb{Z}_4$ and $S = \{ 1,2 \}$,
we obtain the group lattice drawn in Fig.~\ref{fig:z4}.
\begin{figure}
\begin{center}
\includegraphics[scale=.7]{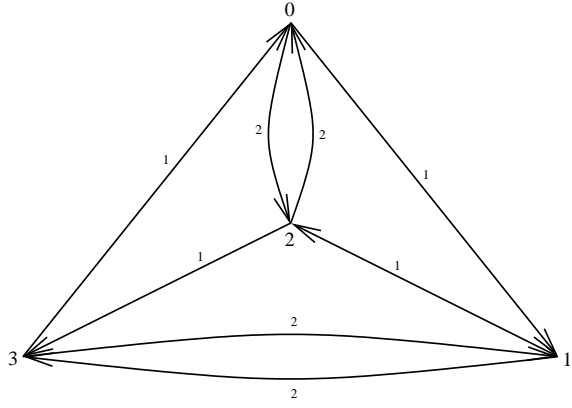}
\caption{The group lattice of $\mathbb{Z}_4$ with
$S = \{1,2\}$.}\label{fig:z4}
\end{center}
\end{figure}
\end{example}

\begin{example}
\label{ex:G=S3}
The permutation group ${\cal S}_3$ has the 6 elements
\bez
  e \, , \quad (12),(13),(23) \, , \quad (123),(132)
\eez
grouped into conjugacy classes.
Choosing $S = \{ (12),(13),(23) \}$, we have three
left-invariant 1-forms $\theta^{(12)}, \theta^{(13)}, \theta^{(23)}$.
The corresponding digraph is drawn on the left-hand side of
Fig.~\ref{fig:threes3}.
Here a line represents a double arrow.
\begin{figure}
\begin{center}
\includegraphics[scale=.6]{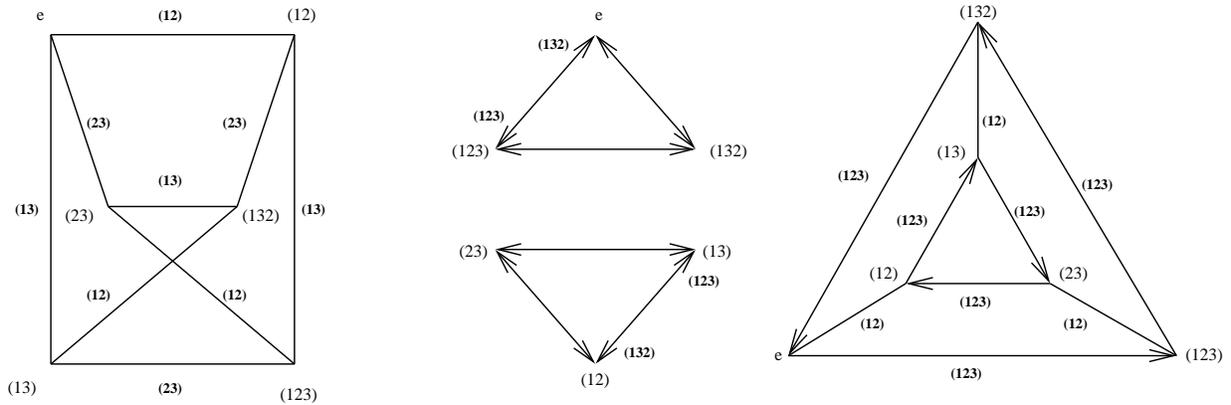}
\caption{Digraphs corresponding to the three different choices
$\{ (12),(13),(23) \}$, $\{ (123),(132) \}$ and $\{ (12),(123) \}$
of $S \subset {\cal S}_3$.}\label{fig:threes3}
\end{center}
\end{figure}

If we choose $S = \{ (123),(132) \}$, then $S$ does not generate ${\cal S}_3$
and the digraph is disconnected. The two parts are drawn in the middle
of Fig.~\ref{fig:threes3}. Since $S_e$ is a
subgroup, according to Lemma~\ref{lemma:subgroup-univdc} we have the
universal first order differential calculus on
the two disjoint parts of ${\cal S}_3$ in this case.

Another choice is $S = \{ (12),(123) \}$. The corresponding digraph
is shown on the right-hand side of Fig.~\ref{fig:threes3} (see also
Refs. \cite{Bollo98,MKS76}).
\end{example}

We call a group lattice {\em bicovariant} if $\ad(S) S \subset S$.
The significance of this definition will be made clear in
section~\ref{sec:diffmaps}.
Our previous examples of group lattices are indeed bicovariant, except
for $({\cal S}_3, S = \{ (12),(123) \})$. Since
$S$ is assumed to be a {\em finite} set, we have the following result.

\begin{lemma}
\label{lemma:adS}
\be
    \ad(g) S \subset S \quad \Rightarrow \quad
    \ad(g^{-1}) S \subset S  \; .
\ee
\end{lemma}
{\bf Proof:} By assumption, $\ad(g)$ is a map $S \rightarrow S$
which is clearly injective. Since $S$ is a finite set, it is
then also surjective. As a consequence,
$\ad(g^{-1}) S = \ad(g)^{-1} S = S$.
\hfill $\blacksquare$
\vskip.2cm

\begin{example}
\label{ex:A5}
Let $G = A_5$, the alternating group consisting of the even permutations
of five objects. It is generated by the two permutations
$a = (12345)$ and $b =(12)(34)$ which satisfy $a^5 = e$, $b^2 = e$ and
$(ab)^3 = e$. Let $S = \{ a, a^{-1}, b \}$. Then the group lattice
is a truncated icosahedron, obtained from the icosahedron by replacing
each of the 12 sites by a pentagon. The result is a group lattice
structure for the $C_{60}$ Fullerene \cite{Samu92}. This group
lattice is not bicovariant.
\end{example}

\begin{figure}
\begin{center}
\includegraphics[scale=.8]{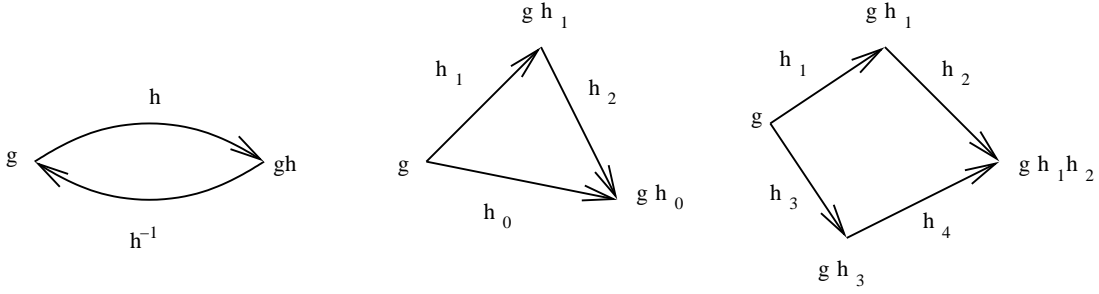}
\caption{Group lattice parts corresponding to a biangle, a triangle
and a quadrangle, respectively.}\label{fig:angle}
\end{center}
\end{figure}

In the following we refer to a pair of elements $h_1,h_2 \in S$ such
that $h_1 h_2 = e$ as a ``biangle'', to a triple $h_0,h_1,h_2 \in S$
such that $h_1 h_2 = h_0$ as a ``triangle'' and to a quadruple
of elements $h_1,h_2,h_3,h_4 \in S$ such that
$h_1 h_2 = h_3 h_4 \not \in S_e$ as a ``quadrangle''
(see Fig.~\ref{fig:angle}).\cite{combin}
In particular, each pair $h_1, h_2$ of commuting elements
of $S$ with $h_1 h_2 \not \in S_e$ determines a quadrangle.

\section{Differentiable maps between group lattices}
\label{sec:diffmaps}
\setcounter{equation}{0}
Let $(G_i,S_i)$, $i=1,2$, be two group lattices and $\phi : G_1 \to
G_2$ a map between them. The latter induces an algebra homomorphism
$\phi^\ast : \A_2 \to \A_1$ where $\phi^\ast f_2 = f_2 \circ \phi$.
In particular,
\be
   \phi^\ast \, e^{g_2} = e^{\phi^{-1} \{g_2\} }
\ee
where we introduced the notation
\be
    e^K := \sum_{g \in K} e^g  \label{e^K}
\ee
for $K \subset G$, and $e^\emptyset := 0$.
The following result shows that every homomorphism between algebras
of functions on group lattices is realized by a pull-back map
(see also Ref. \cite{Dima+TzanIII}).

\begin{theorem}
\label{theor:hom}
If $\Phi : \A_2 \to \A_1$ is an algebra homomorphism,
then there is a map $\phi : G_1 \to G_2$, such that $\Phi = \phi^\ast$.
\end{theorem}

\noindent
{\bf Proof:}
If $f \in \A_1$ is such that $f^2 = f$, then $f = e^K$ for some
$K \subset G_1$. In fact, since $f = \sum_{g_1 \in G_1} f(g_1) \,
e^{g_1}$, we find $f(g_1) (f(g_1)-1) = 0$ for all $g_1 \in G_1$,
so that $f(g_1) \in \{0,1\}$.
Hence $f = \sum_{g_1 \in K} e^{g_1}$ with $K = \{g_1 \in G_1 \,| \,
f(g_1) = 1 \}$.
 From $e^{g_2} e^{g'_2} = \dl^{g_2,g'_2} \, e^{g_2}$ in $\A_2$ we find
$\Phi(e^{g_2}) \Phi(e^{g'_2}) = \dl^{g_2,g'_2} \, \Phi(e^{g_2})$. Hence
$\Phi(e^{g_2}) = e^{K_{g_2}}$ for some $K_{g_2} \subset G_1$.
Furthermore, from $\Phi(e^{g_2}) \Phi(e^{g'_2}) = 0$ for $g_2\neq g'_2$ we
infer $K_{g_2} \cap K_{g'_2} = \emptyset$ and from
$\Phi(\mathbf{1}_2) = \mathbf{1}_1$ we obtain
$\bigcup_{g_2 \in G_2} K_{g_2} = G_1$.
Hence we have a partition of $G_1$. Now we define $\phi : G_1 \to G_2$ by
setting $\phi(g_1) = g_2$ for all $g_1 \in K_{g_2}$. Then $\phi$ is
well defined and $\phi^\ast (e^{g_2}) = e^{K_{g_2}} = \Phi(e^{g_2})$.
\hfill $\blacksquare$
\vskip.1cm

Now we try to extend $\phi^\ast$ to 1-forms requiring
\be
  \phi^\ast (f \, \d_2 f') = (\phi^\ast f) \, \d_1(\phi^\ast f') \, .
\ee
However, this is not well defined unless it is guaranteed that
the right side vanishes whenever the left side vanishes. By linearity,
it is sufficient to consider
\be
  \phi^\ast (e^{g_2} \, \d_2 e^{g'_2}) = e^{\phi^{-1} \{g_2\} }
    \, \d_1 e^{\phi^{-1} \{g'_2\} }
\ee
for all $g_2, g_2' \in G_2$. The consistency condition now takes the
form $\phi^{-1}{\cal I}_2 \subset {\cal I}_1$, which is equivalent to
\be
  g_1^{-1} g'_1 \in S_1 \quad \Longrightarrow \quad
  \phi(g_1)^{-1} \phi(g'_1) \in S_2 \cup \{e_2\} \; .
            \label{phi_diff}
\ee
This means that $\phi$ either sends an arrow at a site to an arrow
at the image site or deletes it, but $\phi$ cannot ``create'' an arrow.
A map with this property will be called {\em differentiable} (see
also Ref. \cite{DMHV95}). In this case we have more generally
$\phi^\ast (f \alpha) = (\phi^\ast f)(\phi^\ast \alpha)$
for $f \in \A_2$ and $\alpha \in \O^1_2$.
\vskip.1cm

In order to define a dual of $\phi^\ast$ on vector fields,
$\phi$ has to be a differentiable bijection. Then we set
\be
   \langle \phi_\ast X_1, \alpha_2 \rangle
 = \langle X_1, \phi^\ast \alpha_2 \rangle \circ \phi^{-1}
                  \label{phi_astX_def}
\ee
where $X_1 \in \X_1$ and $\alpha_2 \in \O^1_2$. As a consequence,
we obtain
\be
   \phi_\ast (f \cdot X) = (\phi^{-1 \ast} f) \cdot \phi_\ast X
           \label{phi_ast_fdot}
\ee
and, using (\ref{X_df}), we find
\be
   \phi_\ast X = \phi^{-1 \ast} \, X \, \phi^\ast \, .
       \label{phi_astX}
\ee
\vskip.1cm

In particular, for each $g \in G$ the left translation $L_g : G \to G$
is a differentiable map since if $g'{}^{-1} g'' \in S$,
then also $(gg')^{-1}(gg'')\in S$. The special basis of 1-forms
$\{ \theta^h | h \in S \}$ and the dual basis $\{ \ell_h | h \in S \}$
of vector fields are left-invariant:
\be
  L_g^\ast \theta^h = \theta^h \, , \quad
  L_{g \ast} \ell_h = \ell_h  \qquad (\forall g \in G, \, h \in S) \, .
\ee
Hence the differential calculus of a group lattice is {\em left covariant}\/.
\vskip.1cm

The condition for the right translation $R_g : G \to G$ to be
differentiable is that for $g'{}^{-1}g''\in S$ also
$(g'g)^{-1}(g''g) = g^{-1}(g'{}^{-1}g'')g \in S$. This amounts to
$\ad(g^{-1})h \in S$ for all $h \in S$.
As a consequence of Lemma~\ref{lemma:adS}, differentiability of $R_g$
implies differentiability of $R_{g^{-1}}$ and we obtain
\be
   R_g^\ast \, \theta^h
 = \sum_{g' \in G} (R_g^\ast \, e^{g'}) \, \d \, R_g^\ast \, e^{g'h}
 = \sum_{g'' \in G} e^{g''} \, \d e^{g''ghg^{-1}}
 = \theta^{\ad(g)h}  \; .    \label{R^ast-theta}
\ee
Furthermore,
\be
   R_{g \ast} \ell_h = \ell_{\ad(g^{-1})h}  \, , \qquad
   R_{g^{-1} \ast} \ell_h = \ell_{\ad(g)h}  \; .
       \label{R_ast-ell}
\ee
If $R_g$ and $R_{g'}$ are both differentiable, then also
$R_{gg'}$ and we have $R_{gg'}^\ast = R_g^\ast \circ R_{g'}^\ast$
on 1-forms.
\vskip.2cm

If $R_g$ is differentiable for all $g \in G$,
then the differential calculus is called {\em right covariant}.
A differential calculus which is both left and right covariant
is called {\em bicovariant} \cite{Woro89}.
Bicovariance of a group lattice, as defined in section~\ref{sec:fodc},
is the weaker condition $\ad(h)S \subset S$ (and then also
$\ad(h^{-1})S \subset S$) for all $h \in S$. This means that
for all $h \in S$ the maps $R_h$ and $R_{h^{-1}}$ are
differentiable. If $S$ does not generate $G$, this condition is
indeed weaker than bicovariance of the first order differential
calculus. But then the corresponding digraph is disconnected
(cf Lemma~\ref{lemma:ccu}). So, if $S$ generates $G$, the
bicovariance conditions for the first order differential calculus
and the group lattice coincide.

\section{Higher order differential calculus of a group lattice}
\label{sec:hodc}
\setcounter{equation}{0}
Let $(\O_u,\d_u)$ be the (full) universal differential calculus over
$\A$. Then we have $\O_u = \bigoplus_{r=0}^\infty \O^r_u$ with
$\O^0_u=\A$. Let  ${\cal J}$ be the differential ideal of $\O_u$
generated by ${\cal J}^1$ where $\O^1 = \O^1_u/{\cal J}^1$.
Since ${\cal J}^1$ is homogeneous of
grade 1, the differential ideal ${\cal J}$ is also graded,
${\cal J} = \bigoplus_{r=0}^\infty{\cal J}^r$ with ${\cal J}^0=\{0\}$.
Then $\O = \O_u/{\cal J}$ inherits the grading, i.e.
$\O = \bigoplus_{r=0}^\infty\O^r$ with $\O^0 = \A$.
The projection $\pi_u : \O_u \to \O$ is a
graded algebra homomorphism and we have a differential map
$\d : \O \to \O$ such that $\d \, \pi_u = \pi_u \, \d_u$. It
satisfies $\d^2 = 0$ and has the graded derivation property
(Leibniz rule)
\be
    \d (\omega \, \omega') = (\d \omega) \, \omega'
  + (-1)^r \, \omega \, \d \omega'   \label{Leibniz}
\ee
for all $\omega \in \O^r$ and $\omega' \in \O$. In this section
we explore for group lattices the structure of $\O$ beyond 1-forms.
\vskip.1cm

For $(g,g') \in {\cal I}$ we obtain
$0 = \pi_u \, \d_u (e^g \d_u e^{g'}) = \pi_u (\d_u e^g) \pi_u (\d_u e^{g'})
   = \d e^g \,\d e^{g'}$.
Using (\ref{de_theta}) and introducing $\tilde{g} = g^{-1} g'$, this
results in the 2-form relations
\be
  \sum_{h,h' \in S} \dl^{\tilde{g}}_{hh'} \, \theta^h \, \theta^{h'} = 0
     \qquad \forall \tilde{g} \not \in S_e  \; .
         \label{2form_rels}
\ee
If $S_e$ is a subgroup of $G$, there are no such conditions. In
this case, the group lattice is disconnected with
components the left cosets of $S_e$ in $G$ and with the universal
differential calculus on each component (see Lemma~\ref{lemma:subgroup-univdc}).
If $S_e$ is not a subgroup, then there are elements $h,h' \in S$
such that $h h' \not\in S_e$ and therefore non-trivial relations
of the form (\ref{2form_rels}) appear.
\vskip.2cm

The following well-known result implies that at the level of $r$-forms,
$r>2$, no further relations appear which are not directly taken
into account by the 2-form relations.

\begin{lemma}
\label{lemma:diffideal}
Let $\alpha \in \O_u^1$. The two-sided ideal generated by $\alpha$
and $\d_u \alpha$ is a differential ideal in $\O_u$.
\end{lemma}
{\bf Proof:} This is an immediate consequence of the Leibniz rule
for $\d_u$ and $\d_u^2 = 0$.
\hfill $\blacksquare$
\bigskip

\noindent
{\em Remark.}
If for some $h \in S$ also $h^{-1} \in S$, then the 2-forms
$\theta^h \theta^{h^{-1}}$, $\theta^{h^{-1}} \theta^h$ do not vanish.
As a consequence, we have forms
$\theta^h \theta^{h^{-1}} \theta^h \cdots$
of arbitrarily high order. This could be avoided by setting
$\theta^h \theta^{h^{-1}} = \theta^{h^{-1}} \theta^h = 0$.
However, such a restriction may exclude interesting cases.
For example, one can formulate the Connes and Lott 2-point space
geometry \cite{Conn+Lott90} using $(\mathbb{Z}_2, \{ 1 \})$.
The only non-vanishing 2-form is then $\theta^1 \theta^1$. If we
set this to zero, then every 2-form automatically vanishes, and
thus in particular the curvature of a connection.
Moreover, such 2-form relations imposed ``by hand'' in
general induce higher form relations, which have
to be elaborated and taken into account. The 2-form
$\theta^h \theta^{h^{-1}}$ has the
interesting property that it commutes with all functions.
\hfill $\blacksquare$
\vskip.2cm

Applying $\d$ to $\theta^h = \sum_{g \in G} e^g \,\d e^{gh}$,
using the Leibniz rule for $\d$ and formulas from section~\ref{sec:fodc},
we find
\be
 \d \theta^h = \theta \, \theta^h + \theta^h \, \theta
               - \Delta(\theta^h)    \label{dthetah}
\ee
where
\be
   \Delta (\theta^h)
 = \sum_{h',h'' \in S} \dl^h_{h'h''} \, \theta^{h'} \theta^{h''}
          \label{Delta}
\ee
determines an $\A$-bimodule morphism\cite{Delta-cons}
$\Delta : \O^1 \to \O^2$.
Using (\ref{df_theta}), we obtain\cite{d-notri}
\be
 \d \alpha = \theta \, \alpha + \alpha \, \theta - \Delta(\alpha)
             \label{dalpha_Delta}
\ee
for an arbitrary 1-form $\alpha$.
A special case of this formula is
\be
   \d \theta = 2 \theta^2 - \Delta(\theta) \; .
\ee
As the sum of all basic 2-forms,
$\theta^2 = \sum_{h,h' \in S} \theta^h \theta^{h'}$
comprises all the 2-form relations.
Since $\Delta(\theta)$ contains all ``triangular'' 2-forms, the
difference $\theta^2 - \Delta(\theta)$ consists of the sum of all
nonzero 2-forms of the form $\theta^h \theta^{h'}$ with $h h'= e$.
Introducing
\be
  \Delta^e := \sum_{h \in S_{(0)}} \theta^h \theta^{h^{-1}}
\ee
where $S_{(0)} := \{ h \in S \, | \, h^{-1} \in S \}$, we obtain
\be
    \theta^2 - \Delta(\theta) = \Delta^e  \label{Delta-Deltae}
\ee
and thus
\be
 \d \theta = \theta^2 + \Delta^e = \Delta(\theta) + 2 \Delta^e \, .
             \label{dtheta}
\ee
Let us extend the map $\Delta$ to $\O$ by requiring
\be
     \Delta(f) = 0
\ee
for all $f \in \A$ and
\be
  \Delta(\omega \, \omega') = \Delta(\omega) \, \omega'
  + (-1)^r \omega \, \Delta(\omega')
                 \label{Delta-Leibniz}
\ee
for all $\omega \in \O^r$ and $\omega' \in \O$.
This is just the (graded) Leibniz rule, hence $\Delta$ is a graded
derivation.

\begin{lemma}
\label{lemma:domega}
\be
     \d \omega = [ \theta , \omega ] - \Delta( \omega )
                    \qquad \forall \, \omega \in \O
                    \label{domega_Delta}
\ee
where $[ \; , \; ]$ is the graded commutator.
\end{lemma}
{\bf Proof:} We use induction on the grade $r$ of forms $\omega \in \O^r$.
For 0-forms the formula is just (\ref{df_theta}), for
1-forms it coincides with (\ref{dalpha_Delta}). Let us now
assume that it holds for forms of grade lower than $r$. For
$\psi \in \O^k$, $k < r$, and $\omega \in \O^{<r}$ we then obtain
\bez
     \d( \psi \omega)
 &=& (\d \psi) \, \omega + (-1)^r \psi \, \d \omega \\
 &=& \Big( [ \theta, \psi] - \Delta(\psi) \Big) \, \omega
     +(-1)^r \, \psi \, \Big( [\theta,\omega] - \Delta(\omega) \Big) \\
 &=& [\theta, \psi \, \omega] - \Delta(\psi \, \omega)
\eez
using the Leibniz rules for $\d$ and $\Delta$.
\hfill $\blacksquare$
\bigskip

Iterated application of (\ref{Delta-Leibniz}) leads to
\be
     \Delta( \theta^{h_1} \cdots \theta^{h_r} )
 &=& \Delta(\theta^{h_1}) \, \theta^{h_1} \cdots \theta^{h_r}
  - \theta^{h_1} \, \Delta(\theta^{h_2}) \, \theta^{h_3} \cdots \theta^{h_r}
     + \ldots  \nonumber \\
 & & + (-1)^{r-1} \, \theta^{h_1} \cdots \theta^{h_{r-1}} \,
     \Delta(\theta^{h_r})    \; .   \label{Delta(omega)}
\ee
Furthermore,
\be
 0 &=& \d^2 \omega = [ \theta , \d \omega ] - \Delta( \d \omega )
               \nonumber \\
   &=& [ \theta , [ \theta , \omega ] ] - [ \theta , \Delta( \omega )]
       - \Delta( [ \theta , \omega ] ) + \Delta^2(\omega) \nonumber \\
   &=& [ \theta^2 - \Delta( \theta ) , \omega ] + \Delta^2(\omega)
\ee
shows that
\be
    \Delta^2 (\omega) = - [\Delta^e , \omega] \; .
\ee
Acting with $\Delta$ on (\ref{Delta-Deltae}), using
(\ref{Delta-Leibniz}) and the last identity, we deduce
\be
    \Delta( \Delta^e ) = 0  \; .  \label{Delta(Deltae)}
\ee
\vskip.1cm
\noindent
{\em Remark.} The cohomology of the universal differential calculus
is always trivial. But this does not hold for its reductions, in general.
For example, for $m > 2$, the group lattice $(\mathbb{Z}_m, \{ 1 \})$
has nontrivial cohomology. There is only a single basis 1-form $\theta^1$
and the 2-form relations enforce $(\theta^1)^2 = 0$ so that there are
no non-vanishing 2-forms. In particular, $\d \theta^1 = 0$. But we
have seen in example~\ref{ex:G=Zm,S=1} that $\theta^1$ is not exact.
The cohomology of the group lattice $(\mathbb{Z}_4, \{ 1,2 \})$, for
example, is trivial.
\hfill $\blacksquare$

\subsection{Action of differentiable maps on forms}
According to section~\ref{sec:diffmaps}, a map $\phi : G \rightarrow G$
is differentiable (with respect to a group lattice structure determined
by a choice $S \subset G$) if the pull-back $\phi^\ast$ extends from
$\A$ to the first order differential calculus, i.e. it also acts on
$\O^1$ as an $\A$-bimodule homomorphism and satisfies
$\phi^\ast (\d f) = \d (\phi^\ast f)$. Moreover, we can extend it to
the whole of $\O$ as an algebra homomorphism via
\be
  \phi^\ast (\omega \, \omega') = (\phi^\ast \omega) (\phi^\ast \omega') \; .
\ee

\begin{lemma}
\label{lemma:phiastd}
For a differentiable map $\phi : G \rightarrow G$ we have
\be
    \phi^\ast \circ \d = \d \circ \phi^\ast  \qquad (\mbox{on } \O ) \; .
        \label{phiast_d}
\ee
\end{lemma}
{\bf Proof:} Since $\phi$ is differentiable, the formula holds on
$0$-forms. If it holds on $r$-forms, then
\bez
     \phi^\ast \d( f \d \omega)
 &=& \phi^\ast (\d f \, \d \omega)
  = (\phi^\ast \d f) \, \phi^\ast \d \omega
  = (\d \phi^\ast f) \, \d \phi^\ast \omega  \\
 &=& \d [ (\phi^\ast f) \, \d \phi^\ast \omega ]
  = \d \, \phi^\ast (f \, \d \omega) \, .
\eez
Since every $(r+1)$-form can be written as a sum of terms like $f \, \d \omega$
with $f \in \A$ and $\omega \in\O^r$, the formula holds for $(r+1)$-forms
and thus on $\O$ by induction.
\hfill $\blacksquare$
\vskip.2cm

By definition, a differentiable map $\phi : G \rightarrow G$
preserves the 1-form relations. Since $\phi^\ast$ commutes with $\d$,
it also preserves the 2-form relations.

\begin{lemma}
\label{lemma:phiasttheta}
For a differentiable bijection $\phi : G \rightarrow G$ we have
\be
     \phi^\ast \theta &=& \theta   \label{phiast_theta}  \\
     \Delta \circ \phi^\ast &=& \phi^\ast \circ \Delta
                    \label{Delta_phiast}
\ee
\end{lemma}
{\bf Proof:} First we note that (\ref{theta_def}) can be written as
\bez
   \theta
 = \sum_{(g,g') \not\in {\cal I}} e^g \, \d e^{g'}
   - \sum_{g \in G} e^g \, \d e^g
 = \sum_{g,g' \in G} e^g \, \d e^{g'}
   - \sum_{g \in G} e^g \, \d e^g \; .
\eez
Then, using $\phi^\ast e^g = e^{\phi^{-1}(g)}$, we find
\bez
   \phi^\ast \theta
 = \sum_{g,g' \in G} e^{\phi^{-1}(g)} \, \d e^{\phi^{-1}(g')}
   - \sum_{g \in G} e^{\phi^{-1}(g)} \, \d e^{\phi^{-1}(g)}
 = \theta
\eez
since $\phi$ is bijective. The second assertion now follows from
\bez
    [\phi^\ast \theta, \phi^\ast \omega] - \phi^\ast \Delta(\omega)
  = \phi^\ast \d \omega
  = \d \phi^\ast \omega
  = [\theta, \phi^\ast \omega] - \Delta(\phi^\ast \omega) \; .
\eez
\hfill $\blacksquare$

\subsection{The structure of the space of 2-forms}
\label{subsec:str2f}
Let $S_{(1)}$ denote the subset of $S$, the elements of which
can be written as products of two other elements of $S$, i.e.
$S_{(1)} = S^2 \cap S$ where $S^2 = \{ hh' | \, h,h' \in S \}$.
Furthermore, let $S_{(2)}$ be the set of
elements of $G$ which do not belong to $S_e$, but can be written
as a product $h h'$ for some $h,h' \in S$.
Hence $S_{(2)} = S^2 \setminus S_e$.
Since for every element of $S_{(2)}$ there is a 2-form relation,
the number of independent 2-forms is $|S|^2-|S_{(2)}|$.
Now we have a decomposition
$S \times S = \{ (h,h^{-1}) \, | \, h \in S_{(0)} \}
   \cup \{ (h,h') \, | \, hh' \in S_{(1)} \}
   \cup \{ (h,h') \, | \, hh' \in S_{(2)} \}$
which defines a direct sum decomposition of $\O^2$.
Introducing projections
\be
     p_{(e)}(\theta^{h_1} \theta^{h_2})
 &=& \dl^e_{h_1h_2} \, \theta^{h_1} \theta^{h_2}  \label{e_proj} \\
     p_{(h)}(\theta^{h_1} \theta^{h_2})
 &=& \dl^h_{h_1h_2} \, \theta^{h_1} \theta^{h_2}  \qquad (h \in S_{(1)})
                                                  \label{h_proj} \\
     p_{(g)}(\theta^{h_1} \theta^{h_2})
 &=& \dl^g_{h_1h_2} \, \theta^{h_1} \theta^{h_2} \qquad (g \in S_{(2)})
                                                  \label{g_proj}
\ee
which extend to left $\A$-module homomorphisms
$p_{(e)}, p_{(h)}, p_{(g)} : \O^2 \to \O^2$,
every 2-form $\psi \in \O^2$ can be
decomposed with the help of the identity
\be
 \psi = ( p_{(e)} + \sum_{h \in S_{(1)}} p_{(h)}
        + \sum_{g \in S_{(2)}} p_{(g)} ) \, \psi  \; .
        \label{psi_decomp}
\ee
The three parts of this decomposition correspond, respectively,
to biangles, triangles and quadrangles, which we introduced
in section~\ref{sec:fodc}.
\vskip.2cm

A relation between elements of $S$ which leads to a 2-form relation
has the form $h_1 h'_1 = h_2 h'_2 = \cdots = h_k h'_k \not\in S_e$.
The latter then implies the 2-form relation
\be
  \theta^{h_1} \theta^{h'_1} + \theta^{h_2} \theta^{h'_2}
 + \cdots + \theta^{h_k} \theta^{h'_k} = 0 \, .
        \label{2form_chain}
\ee
Let us now assume that $(G,S)$ is bicovariant.
Given $h_1, h_2 \in S$ with
$h_1 h_2 \not\in S_e$, we then obtain a chain
$\ldots = h_0 h_1 = h_1 h_2 = h_2 h_3 = \ldots$
where $h_0 = \ad(h_1) h_2$ and $h_3 = \ad(h_2^{-1}) h_1$, and so
forth. Since $S$ is assumed to be finite, only a finite part of
the chain contains pairwise different members. This means that
the chain must actually consist of ``cycles'', i.e. subchains of the form
$h_1 h_2 = h_2 h_3 = \cdots = h_{r-1} h_r = h_r h_1$.
A relation like $\theta^h \theta^{h'} = 0$, consisting of a
single term, is only possible if $h' = h$ and $h^2 \not\in S_e$.
\vskip.2cm

\begin{example}
\label{ex:S3-2forms}
For the permutation group ${\cal S}_3$ and $S = \{ (12),(13),(23) \}$
(see example~\ref{ex:G=S3}) we have
$S_{(0)} = S$ (since $(ij)^2 = e$), $S_{(1)} = \emptyset$ and
$S_{(2)} = \{ (123),(132) \}$. As a consequence of the cycles
$(12)(13) = (13)(23) = (23)(12) = (123)$ and
$(12)(23) = (23)(13) = (13)(12) = (132)$ the
three basic 1-forms $\theta^{(12)}, \theta^{(13)}, \theta^{(23)}$
have to satisfy the two 2-form relations
\bez
    \theta^{(12)} \theta^{(13)} + \theta^{(13)} \theta^{(23)}
  + \theta^{(23)} \theta^{(12)} = 0 \, ,
            \qquad
    \theta^{(12)} \theta^{(23)} + \theta^{(23)} \theta^{(13)}
  + \theta^{(13)} \theta^{(12)} = 0 \, .
\eez
Hence there are $3^2-2=7$ independent 2-forms:
$\theta^{(12)} \theta^{(12)}$, $\theta^{(13)} \theta^{(13)}$,
$\theta^{(23)} \theta^{(23)}$ and, say, $\theta^{(13)} \theta^{(23)}$, $\theta^{(23)} \theta^{(12)}$,
$\theta^{(12)} \theta^{(23)}$, $\theta^{(23)} \theta^{(13)}$.

If we choose $S = \{(123),(132)\}$, then $S_e$ is a subgroup and
we have the universal calculus on the two cosets of $S_e$ in
${\cal S}_3$. Then there are no 2-form relations.
\end{example}

\begin{example}
\label{ex:A4-2forms}
The alternating group $A_4$ has the following elements,
\bez
  e \, , \quad  (123),(243), (134),(142) \, , \quad
 (132),(234),(143),(124) \, , \quad (12)(34), (13)(24), (14)(23)
\eez
grouped into conjugacy classes. Choosing
$S = \{(123),(243),(134),(142)\}$, the group lattice is connected.
As a consequence of
\bez
 (123)(134) = (134)(243) = (243)(123) &=& (124) = (142)^2   \\
 (123)(243) = (243)(142) = (142)(123) &=& (143) = (134)^2   \\
 (123)(142) = (142)(134) = (134)(123) &=& (234) = (243)^2   \\
 (134)(142) = (142)(243) = (243)(134) &=& (132) = (123)^2
\eez
we obtain four 2-form relations, so there are twelve independent
2-forms. Note that in this example there are two different
cycles for each of the elements $(124),(143),(234),(132)$
of $S_{(2)}$.
\end{example}

\noindent
{\em Remark.}
For a bicovariant differential calculus a bimodule isomorphism
$\sigma : \O^1 \oA \O^1 \rightarrow \O^1 \oA \O^1$ exists
\cite{Woro89,BDMHS96} such that
\be
   \sigma(\theta^{h_1} \oA \theta^{h_2})
 = \theta^{\ad(h_1) h_2} \oA \theta^{h_1}
 = \theta^{h_0} \oA \theta^{h_1}
\ee
with inverse
\be
   \sigma^{-1}(\theta^{h_1} \oA \theta^{h_2})
 = \theta^{h_2} \oA \theta^{\ad(h_2^{-1}) h_1}
 = \theta^{h_2} \oA \theta^{h_3} \; .
\ee
These formulas show that the 2-form relations, and moreover each
cycle, is invariant under $\sigma$.
Woronowicz \cite{Woro89} introduced the wedge product
\be
   \theta^h \wedge \theta^{h'} = {1 \over 2} (\mbox{id} - \sigma)
    (\theta^h \oA \theta^{h'})    \; .
\ee
A particular consequence is
\be
   \theta^h \wedge \theta^h = 0  \; .
\ee
Furthermore, for every cycle there is a 2-form relation.
For example,
\be
   \theta^{h_1} \wedge \theta^{h_2} + \cdots
 + \theta^{h_r} \wedge \theta^{h_1} = 0
\ee
for the cycle $h_1 h_2 = h_2 h_3 = \cdots = h_r h_1$.
This means that the Woronowicz wedge product refines our
2-form relations by decoupling cycles
belonging to the same $g \in S_{(2)}$ and imposing a separate
2-form relation for each cycle.
In the example of the alternating group $A_4$, this yields eight
conditions from the previous four, e.g. instead of
\be
 \theta^{(123)} \theta^{(134)} + \theta^{(134)} \theta^{(243)}
 + \theta^{(243)} \theta^{(123)} + \theta^{(142)}\theta^{(142)} = 0
\ee
we obtain
\be
     \theta^{(123)} \wedge\theta^{(134)}
   + \theta^{(134)} \wedge \theta^{(243)}
   + \theta^{(243)} \wedge \theta^{(123)}
 = 0
 = \theta^{(142)} \wedge \theta^{(142)} \; .
\ee
\hfill $\blacksquare$
\vskip.2cm

\begin{example}
\label{ex:S4-2forms}
Besides the unit element $e$, the group ${\cal S}_4$ of
permutations of four objects has the following 23 elements
\bez
  && (12),(13),(14),(23),(24),(34)   \\
  && (123),(124),(132),(134),(142),(143),(234),(243)  \\
  && (12)(34),(13)(24),(14)(23)  \\
  && (1234),(1243),(1324),(1342),(1423),(1432)
\eez
grouped into conjugacy classes. Choosing
$S = \{ (12),(13),(14),(23),(24),(34) \}$, we find $S_{(0)} = S$,
$S_{(1)} = \emptyset$ and
\bez
 S_{(2)} = \{ (123),(132),(124),(142),(134),(143),(234),(243),
              (12)(34),(13)(24),(14)(23) \} \, .
\eez
Hence there are eleven 2-form relations and thus $6^2 - 11 = 25$
independent products of two of the 1-forms $\theta^h$, $h \in S$.
Six of them are of the form $\theta^{(ij)} \theta^{(ij)}$, $i\neq j$.
These would vanish if we required the Woronowicz wedge product.
\end{example}
\vskip.2cm

Given a 2-form
\be
   \psi = \sum_{h,h' \in S} \psi_{h,h'} \, \theta^h \, \theta^{h'}
\ee
the biangle and triangle coefficient functions $\psi_{h,h'}$ are uniquely
determined, but there is an ambiguity in the quadrangle coefficients
due to the 2-form relations (\ref{2form_rels}).
As a consequence of the latter, writing
\be
  \psi_{(g)} = p_{(g)} \psi
  = \sum_{h,h' \in S} \check{\psi}_{(g) \, h,h'} \, \theta^h \, \theta^{h'}
\ee
for $g \in S_{(2)}$, there is a freedom of gauge transformations
$\check{\psi}_{(g) \, h,h'} \mapsto \check{\psi}_{(g) \, h,h'}
+ \Psi_{(g)} \, \delta^g_{hh'}$
with an arbitrary function $\Psi_{(g)}$ on $G$.
The differences
\be
    \psi_{(g) \, h, h'; \hat{h}, \hat{h}'}
  := \check{\psi}_{(g) \, h,h'} - \check{\psi}_{(g) \, \hat{h},\hat{h}'}
\ee
are gauge invariant for all pairs $h,h'$ and $\hat{h},\hat{h}'$
with $\hat{h}\hat{h}' = hh' = g$. As a consequence,
the \emph{quadrangle components} of $\psi$ (with $hh'=g$) defined
in a symmetric way by
\be
    \psi_{(g) \, h,h'}
 := \sum_{\hat{h},\hat{h}'} \dl^g_{\hat{h} \hat{h}'}
    \, \psi_{(g) \, h,h'; \hat{h},\hat{h}'}
  = |g| \, \check{\psi}_{h,h'} - \sum_{\hat{h},\hat{h}'}
    \dl^g_{\hat{h}\hat{h}'} \, \check{\psi}_{\hat{h},\hat{h}'}
    \label{psi_quadr_comp}
\ee
with $|g| := \sum_{h,h'} \delta^g_{hh'}$
are independent of the choice of the coefficient functions
$\check{\psi}_{(g) \, h,h'}$ (from their gauge equivalence class).
They satisfy $\sum_{h,h'} \dl^g_{hh'} \, \psi_{(g)\, h,h'} = 0$ and
\be
  \psi_{(g)} = {1 \over |g|} \, \sum_{h,h' \in S} \dl^g_{hh'}
   \, \psi_{(g) \, h,h'} \, \theta^h \, \theta^{h'} \; .
\ee
The equation $\psi_{(g)} = 0$ for a 2-form $\psi$ is equivalent to the
vanishing of all the differences $\psi_{(g) \, h, h'; \hat{h}, \hat{h}'}$
where $hh' = \hat{h}\hat{h}' = g$.
\vskip.2cm

\begin{example}
\label{ex:S3quadcomp}
Consider $G = {\cal S}_3$ with $S = \{(12),(13),(23)\}$, see
example~\ref{ex:S3-2forms}. A 2-form
\be
     \psi
 &=& \sum_{(ij),(kl) \in S} \psi_{(ij),(kl)} \,
     \theta^{(ij)} \, \theta^{(kl)}
  = \sum_{(ij) \in S} \psi_{(e)\,(ij),(ij)} \,
     \theta^{(ij)} \, \theta^{(ij)}   \nonumber \\
 & & + \sum_{(ij),(kl) \in S \atop (ij)(kl)=(123)}
       \check{\psi}_{((123)) \, (ij),(kl)} \, \theta^{(ij)} \, \theta^{(kl)}
     + \sum_{(ij),(kl) \in S \atop (ij)(kl)=(132)}
       \check{\psi}_{((132)) \, (ij),(kl)} \, \theta^{(ij)} \, \theta^{(kl)}
\ee
then has biangle components $\psi_{(e) \, (ij),(ij)}$, $(ij) \in S$.
The quadrangle components are
\be
     \psi_{((123)) \, (12),(13)}
 &=& 2 \, \check{\psi}_{((123)) \, (12),(13)}
     - \check{\psi}_{((123)) \, (13),(23)} - \check{\psi}_{((123)) \, (23),(12)}
                \nonumber \\
     \psi_{((123)) \, (13),(23)}
 &=& 2 \, \check{\psi}_{((123)) \, (13),(23)}
     - \check{\psi}_{((123)) \, (12),(13)} - \check{\psi}_{((123)) \, (23),(12)}
                \nonumber \\
     \psi_{((123)) \, (23),(12)}
 &=& 2 \, \check{\psi}_{((123)) \, (23),(12)}
     - \check{\psi}_{((123)) \, (12),(13)} - \check{\psi}_{((123)) \, (13),(23)}
\ee
and similar expressions for $\psi_{((132)) \, (ij),(kl)}$.
\end{example}

\section{Discrete and basic vector fields}
\label{sec:discr_vf}
\setcounter{equation}{0}
A vector field is by definition an expression of the form
$X = \sum_{h \in S} X^h \cdot \ell_h$ with $X^h \in \A$.
In this section we explore the properties of special classes
of vector fields.

\subsection{Discrete vector fields}
A vector field will be called {\em discrete} if it has
the property
\be
   X(ff') = (Xf) f' + f (Xf') + (X f)(Xf')
   \qquad \forall \, f,f' \in \A  \; . \label{discr_vf}
\ee
As a consequence, its components satisfy
\be
   X^h \, X^{h'} = \dl^{h,h'} \, X^h  \qquad \forall \, h,h' \in S
    \, .  \label{XX}
\ee
This implies that, at every site $g$, the components $X^h(g)$
all vanish except for at most one component which must then be equal
to 1. In particular, the vector fields $\ell_h$ satisfy
(\ref{discr_vf}) and are therefore discrete.
\vskip.1cm

Discrete vector fields are precisely those vector fields which
describe a deterministic motion on a group lattice in the following
way, where $S$ specifies the possible ``directions''. A ``particle''
moving on a group lattice stops if it reaches a site $g$ where
$X^h(g)=0$ for all $h \in S$. It moves further to $gh$ if
$X^h(g)=1$ for some $h \in S$.
\vskip.1cm

In the following, a visualization is helpful.
If $X^h(g) = 1$, we assign an ``$X$-arrow''
to the site $g$ pointing to the site $gh$ in the
group lattice.
\vskip.2cm

\noindent
{\em Remark.}
An important generalization of discrete vector fields is given
by vector fields $P = \sum_{h \in S} P^h \cdot \ell_h$ satisfying
$P^h \geq 0$ and $\sum_{h \in S} P^h \leq \mathbf{1}$.
In the context of random walks, $P^h(g)$ may be interpreted as
the probability for a move from $g$ to $gh$. Then
$P^e(g) := 1 - \sum_{h \in S} P^h(g)$ is the
probability for a rest at the site $g$.
See also Refs. \cite{Dima+TzanIII,Salo01}.
\hfill $\blacksquare$

\vskip.2cm
It is convenient to introduce $X^e$ such that $X^e(g) = 1$
if $X^h(g) = 0$ for all $h \in S$, and $X^e(g) = 0$ otherwise.
Then we have the useful formula
\be
   (I+X) f = \sum_{h \in S_e} X^h \, R^\ast_h f
              \label{I+X}
\ee
where $I$ is the identity on $\A$.

\begin{lemma}
\label{lemma:dvf-phi}
If $X$ is a discrete vector field, then $I+X$ is an endomorphism of $\A$
and there is a map $\phi_X : G \to G$
such that
\be
    I+X = \phi_X^\ast   \qquad \mbox{on } \A \; . \label{I+X=phi_X}
\ee
\end{lemma}
{\bf Proof:} Using (\ref{discr_vf}), it is easily verified
that $I+X$ is an algebra homomorphism $\A \rightarrow \A$.
Then theorem~\ref{theor:hom} ensures the existence of a map
$\phi_X : G \to G$ with $I+X = \phi_X^\ast$.
\hfill $\blacksquare$
\bigskip

A more concrete description of the map $\phi_X$ is obtained below.
Since for each $g \in G$ there is precisely one $h \in S_e$
with $X^h(g) = 1$, a discrete vector field $X$ determines
a map $s_X : G \to S_e$. Then
\be
  X = \sum_{h \in S} X^h \cdot \ell_h
    = \sum_{g \in G} e^g \cdot \ell_{s_X(g)}
    =: \ell_{s_X}          \label{X_s}
\ee
(where $\ell_e = 0$).
Conversely, every map $s : G \to S_e$ defines a discrete
vector field via the last formula. In fact, this correspondence
between discrete vector fields and maps $G \to S_e$ is
easily seen to be bijective.

\vskip.2cm
Let us now define
\be
   \phi_X(g) := g \, s_X(g)  \; .  \label{phi_X-s}
\ee
Then we obtain
\be
   (\phi_X^\ast f)(g) = f( \phi_X(g)) = f(g \, s_X(g))
   = \sum_{h \in S_e} X^h(g) \, (R_h^\ast f)(g)
\ee
so that
\be
   \phi_X^\ast = \sum_{h \in S_e} X^h \, R^\ast_h
     \qquad \mbox{on $\A$} \, .   \label{phi_X^ast}
\ee
Now (\ref{I+X}) shows that $I+X = \phi_X^\ast$
on functions, in accordance with lemma~\ref{lemma:dvf-phi}.
\vskip.2cm

 For a map $\phi : G \rightarrow G$ the expression
$\phi^\ast - I$ is in general not a vector field. For example,
since $(X^2 f)(g)$ in general also depends on the values of $f$
at over-next neighbors, $\phi_X^2 - I = 2 \, X + X^2$ is not
a vector field.

\begin{lemma}
\label{lemma:phi-dvf}
For a map $\phi : G \rightarrow G$ the expression
$\phi^\ast-I$ is a discrete vector field if and only if
$(g, \phi(g)) \not \in {\cal I}$ for all $g \in G$.
\end{lemma}
{\bf Proof:} If $(g, \phi(g)) \not \in{\cal I}$, then $g^{-1} \phi(g) \in S_e$
and thus defines a map $s : G \to S_e$ such that $\phi(g) = g \, s(g)$.
This map defines a discrete vector field $X$ such that
$\phi^\ast = \phi_X^\ast = I+X$.
The converse is a simple consequence of (\ref{phi_X-s}).
\hfill $\blacksquare$
\bigskip

Discrete vector fields need not generate differentiable maps.
In fact, since $I + \ell_h = R^\ast_h$, the corresponding condition
for the discrete vector fields $\ell_h$ is $\ad(h^{-1}) h' \in S$
for all $h' \in S$ (see section~\ref{sec:diffmaps}). This condition
is also needed for right covariance of the group lattice differential
calculus, but is weaker than that.

\begin{theorem}
\label{theor:phiXdiff}
For a discrete vector field $X$, the following conditions
are equivalent. \\
(1) $\phi_X$ is differentiable. \\
(2) $(g \, s_X(g))^{-1} \, g' s_X(g') \in S_e$ for all $g,g'$ with
    $g^{-1} g' \in S$. \\
(3) For each discrete vector field $Y$ there is a discrete vector
    field $Z$ such that
    $\phi_Y^\ast \, \phi_X^\ast = \phi_X^\ast \, \phi_Z^\ast$.\cite{intertw}
\end{theorem}

\noindent
{\bf Proof:} Using (\ref{phi_X-s}), the equivalence of
(1) and (2) follows from (\ref{phi_diff}).
With $X = \sum_{h \in S_e} X^h \cdot \ell_h$,
$Y = \sum_{h \in S_e} Y^h \cdot \ell_h$ and
$Z = \sum_{h \in S_e} Z^h \cdot \ell_h$, the formula in (3) reads
\bez
   \sum_{h_1, h \in S_e} Y^h \, (R_h^\ast X^{h_1}) \, R^\ast_{h h_1}
 = \sum_{h_1, h' \in S_e} X^{h_1} \, (R^\ast_{h_1} Z^{h'}) \, R^\ast_{h_1 h'}
   \; .
\eez
Hence, for all $g,g' \in G$ we obtain
\bez
   \sum_{h_1, h \in S_e} Y^h(g) \, X^{h_1}(g h) \, \dl^{g'}_{h h_1}
 = \sum_{h_1, h' \in S_e} X^{h_1}(g) \, Z^{h'}(gh_1) \, \dl^{g'}_{h_1h'}
   \; .
\eez
Since $X^h(g) = \dl^h_{s_X(g)}$, this becomes
\bez
 \sum_{h_1, h \in S_e} \dl^h_{s_Y(g)} \, \dl^{h_1}_{s_X(gh)}
   \, \dl^{g'}_{h h_1}
 = \sum_{h_1, h' \in S_e} \dl^{h_1}_{s_X(g)} \, \dl^{h'}_{s_Z(gh_1)}
   \, \dl^{g'}_{h_1 h'}
\eez
and thus
\be
   s_X(g)^{-1} \, s_Y(g) \, s_X(g \, s_Y(g)) = s_Z(g \, s_X(g)) \in S_e \; .
\ee
Since for all $g,g'$ with $g^{-1} g' \in S$ there is a discrete
vector field $Y$ such that $g^{-1} g' = s_Y(g)$, we have shown that
(3) implies (2). Conversely, if (2) holds, then we define
$s_Z(g \, s_X(g))$ by the left-hand side of the above formula.
This determines a discrete vector field $Z$ at all sites except
those which have an outgoing $X$-arrow but no incoming
one. At those sites $g'$, we can choose arbitrary values of $s_Z(g')$.
Then (3) holds.
\hfill $\blacksquare$
\bigskip

\begin{example}
\label{ex:s-maps}
If $\ad(h^{-1}) S \subset S$ for all $h \in S$, then a map $s : G \to S_e$
with the property that for all $g \in G$ we have $s(gh) = s(g)$ for all
$h \in S_e$, solves condition (2) of theorem~\ref{theor:phiXdiff} and
thus defines a discrete vector field $X$ for which $\phi_X$ is
differentiable. But then $X = 0$ or $X = \ell_h$ for some $h \in S$
(on each connected component of the group lattice).

Another example, which trivially satisfies condition (2), is given by
a map with $s(gh) = h^{-1} s(g) h$.
\end{example}

A discrete vector field $X$ generates a discrete flow on $\A$
via $(I+X)^n$, $n=0,1,2,\ldots$ (see also appendix~\ref{sec:int_curv}).
We sometimes refer to $\phi_X^\ast = I + X$ as the {\em flow} of $X$.
If the flow is differentiable, then it extends to $\O$. Moreover,
if $\phi_X$ is also invertible, it induces a map $\phi_{X \ast}$ on
the space $\X$ of vector fields via (\ref{phi_astX_def}).

\subsection{Discrete vector fields with invertible flow}
The following result characterizes discrete vector fields with
invertible flow.

\begin{theorem}
\label{theor:I+X}
Let $X = \sum_{h \in S} X^h \cdot \ell_h$ be a discrete vector field.
The following conditions are equivalent:   \\
(1) $I+X$ is an automorphism of $\A$.      \\
(2) For every $g$ with $X^h(g) = 1$ for some $h \in S$, there is
precisely one $h' \in S$ such that $X^{h'}(g {h'}^{-1}) = 1$.
If $X^h(g)=0$ for all $h \in S$, then also
$X^{h}(g {h}^{-1}) = 0$ for all $h \in S$.\cite{Xarrow} \\
(3) $\sum_{h \in S_e} X^h(g h^{-1}) = 1$ for all $g \in G$.
\end{theorem}

\noindent
{\bf Proof:} We already know that $I+X$ is a homomorphism
(lemma~\ref{lemma:dvf-phi}). Hence $I+X$ is an automorphism
if and only if it is bijective, which means that $(I+X) f' = f$
has a unique solution $f'$ for each $f \in \A$.
This equation reads
\be
   \sum_{h' \in S_e} X^{h'}(g) f'(g h') = f(g)
   \qquad \forall g \in G \, .         \label{I+Xf'f}
\ee
``(1) $\Rightarrow$ (2)'': We assume that $I+X$ is an automorphism.
Let $g$ be such that $X^h(g) = 0$ for all $h \in S$.
Suppose that $X^{h'}(g {h'}^{-1}) = 1$ for some $h' \in S$. Then
(\ref{I+Xf'f}) implies $f'(g) = f(g)$ and also
$f'(g) = f(g {h'}^{-1})$ in contradiction to $I+X$ being
surjective. Hence the second part of condition (2) holds.

Let $g$ be such that $X^h(g) = 1$ for some $h \in S$.
Suppose that $X^{h'}(g {h'}^{-1}) = 0$ for all $h' \in S$. Then
(\ref{I+Xf'f}) places no restrictions on $f'(g)$ which contradicts that
$I+X$ is injective.
Hence $X^{h_g}(g h_g^{-1}) = 1$ for some
$h_g \in S$. Now suppose that there are two different elements
$h_g, h'_g$ with this property.
Then (\ref{I+Xf'f}) implies
$f'(g) = f(g h_g^{-1}) = f(g {h'}_g^{-1})$ which restricts $f$.
This contradicts that $I+X$ is surjective.
Hence the first part of condition (2) holds.
\\
``(2) $\Rightarrow$ (3)'': This is easily verified.
\\
``(3) $\Rightarrow$ (1)'':
Multiplying (\ref{I+Xf'f}) with $X^h(g)$, $h \in S_e$,
and using (\ref{XX}) leads to $X^h(g)( f'(gh) - f(g)) = 0$
for all $g \in G$ and $h \in S_e$.
Hence, for every $g' \in G$ such that $g'{}^{-1} g \in S_e$ we have
\bez
   X^{g'{}^{-1}g}(g') \, [ f'(g) - f(g') ] = 0 \, .
\eez
Condition (3) implies that for each $g$ there is exactly one $g'$
such that $X^{g'{}^{-1}g}(g') = 1$. The above equation then
defines a function $f'$ on $G$. In fact, the latter is given by
\bez
    f'(g) = \sum_{h \in S_e} X^h(g h^{-1}) \, f(gh^{-1}) \, .
\eez
Hence $I+X$ is surjective. Furthermore,
$f=0$ enforces $f'=0$ so that $I+X$ is also injective.
\hfill $\blacksquare$
\bigskip

If $I+X$ is invertible, according to condition
(2) of theorem~\ref{theor:I+X} there is a map
$r_X : G \rightarrow S_e$ such that
$X^h(g h^{-1}) = \delta^{h,r_X(g)}$, i.e.
\be
    R^\ast_{h^{-1}} X^h = \dl^{h,r_X}  \; .  \label{Rast-rX}
\ee
Whereas $s_X$ determines the outgoing $X$-arrow at a site $g$
with $s_X(g) \neq e$, the map $r_X$ determines the corresponding
incoming $X$-arrow.

\begin{lemma}
\label{lemma:XXdiscr-inv}
The components of a discrete vector field $X$ with invertible flow
satisfy
\be
    (R_{h^{-1}}^\ast X^h) \, (R_{{h'}^{-1}}^\ast X^{h'})
  = \dl^{h,h'} \, R_{h^{-1}}^\ast X^h
    \qquad  \forall \, h,h' \in S \; .  \label{XXdiscr-inv}
\ee
\end{lemma}
{\bf Proof:} This follows immediately from (\ref{Rast-rX}).
\hfill $\blacksquare$
\bigskip

As a consequence of (\ref{Rast-rX}), both maps are related by
\be
   s_X( g \, r_X(g)^{-1} ) = r_X(g) \; .  \label{sX-rX}
\ee
Since the incoming $X$-arrow at $g \, s_X(g)$ is the outgoing
$X$-arrow at $g$, also the following relation holds:
\be
   r_X( g \, s_X(g) ) = s_X(g) \; .   \label{rX-sX}
\ee
In particular, these relations imply $r_X(g) = e$ if and only if
$s_X(g) = e$.

\begin{lemma}
\label{lemma:phiX-rX}
Let $X$ be a discrete vector field with invertible flow. Then
\be
    \phi_X^{-1}(g) = g \, r_X(g)^{-1}  \; . \label{phiX-rX}
\ee
\end{lemma}
{\bf Proof:} Using (\ref{phi_X-s}), (\ref{sX-rX}) and (\ref{rX-sX}), we find
\bez
  \phi_X (g \, r_X(g)^{-1}) = g \, r_X(g)^{-1} \, s_X( g \, r_X(g)^{-1} )
  = g \, r_X(g)^{-1} \, r_X(g) &=& g   \\
  \phi_X(g) \, r_X(\phi_X(g))^{-1} =
  g \, s_X(g) \, r_X(g \, s_X(g))^{-1} = g \, s_X(g) \, s_X(g)^{-1} &=& g
\eez
\hfill $\blacksquare$
\bigskip

As a consequence of (\ref{phiX-rX}), on $\A$ we have
\be
   (\phi_X^{-1})^\ast
  = R^\ast_{r_X^{-1}}
  = \sum_{h \in S_e} \dl^{h,r_X} \, R^\ast_{h^{-1}}
  = \sum_{h \in S_e} (R^\ast_{h^{-1}} X^h) \, R^\ast_{h^{-1}} \; .
           \label{phiXast-inv}
\ee

\begin{lemma}
\label{lemma:checkX}
If $I+X$ with a discrete vector field $X$
is invertible on $\A$, its inverse is $I + \check{X}$ with\cite{novec}
$\check{X} = \sum_{h \in S} \delta^{h,r_X} \cdot \ell_{h^{-1}}$.
\end{lemma}
{\bf Proof:} This follows immediately from (\ref{Rast-rX}) and
(\ref{phiXast-inv}).
\hfill $\blacksquare$

\begin{theorem}
\label{theor:phi_Xinv-diff}
Let $X$ be a discrete vector field with invertible flow. Then
$\phi_X$ is differentiable if and only if
$\phi_X^{\ast -1} \, Y \, \phi_X^\ast$ is a discrete
vector field for all discrete vector fields $Y$.
\end{theorem}
{\bf Proof:} According to theorem~\ref{theor:phiXdiff}, $\phi_X$ is
differentiable if and only if for each discrete vector field
$Y$ there is a discrete vector field $Z$ such that
$\phi_Y^\ast \, \phi_X^\ast = \phi_X^\ast \, \phi_Z^\ast$.
Using
\bez
    \phi_X^{\ast -1} \, Y \, \phi_X^\ast
  = \phi_X^{\ast -1} \, (\phi_Y^\ast - I) \, \phi_X^\ast
  = \phi_X^{\ast -1} \, \phi_Y^\ast \, \phi_X^\ast - I
\eez
the last condition translates to
\bez
  \phi_X^{\ast -1} \, Y \, \phi_X^\ast = \phi_Z^\ast - I = Z  \; .
\eez
\hfill $\blacksquare$
\bigskip

\noindent
{\bf Corollary.}
\emph{Let $X$ be a discrete vector field with differentiable and
invertible $\phi_X$.
Then $\phi_{X \ast}$ (defined by (\ref{phi_astX_def})) maps
discrete vector fields to discrete vector fields.}
\vskip.1cm
\noindent
{\bf Proof:} This follows directly from (\ref{phi_astX})
and theorem~\ref{theor:phi_Xinv-diff}.
\hfill $\blacksquare$

\subsection{Another extension of the flow to forms
and vector fields on a bicovariant group lattice}
In this subsection we assume that the group lattice is bicovariant,
so that $R_h$ is differentiable for all $h \in S$
(see section~\ref{sec:diffmaps}). Let $X$ be a discrete vector field.
Then
\be
   R_X \omega := \sum_{h \in S_e} X^h \, R_h^\ast \omega
\ee
directly extends (\ref{phi_X^ast}) from functions to arbitrary
forms.
\vskip.2cm
\noindent
{\em Remark.} Bicovariance does not imply that $\phi_X$
is differentiable. Even if $\phi_X$ is differentiable,
we have in general $\phi_X^\ast \omega \neq R_X \omega$
(see example~\ref{ex:bv-S3} below).
Hence, there are two natural actions on forms, $\phi_X^\ast$ and $R_X$.
They coincide on functions, but differ, in general, on forms.
\hfill     $\blacksquare$
\bigskip

\begin{lemma}
\label{lemma:R_X_inv}
If $X$ is a discrete vector field with invertible flow on a bicovariant
group lattice, then $R_X$ is invertible on $\O$ with
\be
  R_X^{-1} = \sum_{h \in S_e} (R^\ast_{h^{-1}} X^h) \, R^\ast_{h^{-1}}
             \; .    \label{R_X_inv}
\ee
\end{lemma}
{\bf Proof:} We have
\bez
   \sum_{h \in S_e} (R^\ast_{h^{-1}} X^h) \, R^\ast_{h^{-1}} \, R_X
 &=& \sum_{h,h' \in S_e} (R^\ast_{h^{-1}} X^h) \, R^\ast_{h^{-1}} \,
    X^{h'} \, R_{h'}^\ast
  = \sum_{h,h' \in S_e} (R^\ast_{h^{-1}} X^h \, X^{h'})
   \, R^\ast_{h^{-1}h'} \\
 &=& \sum_{h \in S_e} (R^\ast_{h^{-1}} X^h) \, I = I
\eez
where we used (\ref{XX}) and condition (3) of theorem~\ref{theor:I+X}
in the last steps. In a similar way,
$R_X \, \sum_{h \in S_e} (R^\ast_{h^{-1}} X^h) \, R^\ast_{h^{-1}} = I$
is obtained with the help of (\ref{XXdiscr-inv}).
\hfill $\blacksquare$
\bigskip

Assuming that $R_X$ is invertible on $\A$, following (\ref{phi_astX_def})
we define a map $R_{X \ast}$ on vector fields $Y \in \X$ by
\be
   \langle R_{X \ast} Y , \alpha \rangle
 = R_X^{-1} \langle Y , R_X \alpha \rangle  \; .
           \label{R_Xast_def}
\ee

\begin{lemma}
\label{lemma:R_XastY}
Let $X$ be a discrete vector field with invertible flow on a bicovariant
group lattice. Then $R_{X \ast}$ acts on $\X$ as follows,
\be
     R_{X \ast} Y
 &=& \sum_{h \in S_e} (R_{h^{-1}}^\ast X^h) \cdot R_{h \ast} Y  \; .
            \label{R_XastY}
\ee
\end{lemma}
{\bf Proof:} This is obtained from (\ref{R_Xast_def})
using $R_X^{-1} = \phi_X^{\ast-1}$ on functions, (\ref{phiXast-inv}),
(\ref{XX}) and (\ref{phi_astX_def}) applied to the map $R_h$.
\hfill $\blacksquare$
\bigskip

With the help of (\ref{phi_astX}), (\ref{Rast-rX}) and $\ad(h)^{-1} S = S$
for $h \in S_e$, (\ref{R_XastY}) reads
\be
     R_{X \ast} Y
 &=& \sum_{h \in S_e} (R_{h^{-1}}^\ast X^h) \cdot R_{h^{-1}}^\ast Y \, R_h^\ast
  = \sum_{h,h' \in S_e} (R_{h^{-1}}^\ast X^h) \, ( R_{h^{-1}}^\ast Y^{\ad(h)h'} )
    \cdot \ell_{h'}  \label{R_XastY-expl} \\
 &=& \sum_{h,h' \in S_e} \dl^{h,r_X} \, ( R_{h^{-1}}^\ast Y^{\ad(h)h'} )
    \cdot \ell_{h'}  \; .  \nonumber
\ee
If $Y$ is a discrete vector field, further evaluation leads to
\be
     R_{X \ast} Y
  = \sum_{h' \in S_e} \dl^{\ad(r_X)h',s_Y \circ R_{r_X^{-1}}} \cdot \ell_{h'}
  = \ell_{\ad(r_X^{-1}) (s_Y \circ R_{r_X^{-1}})}
  = \sum_{g \in G} e^g \cdot \ell_{\ad(r_X(g)^{-1}) \, s_Y(g \, r_X(g)^{-1})}
    \quad \label{R_XastYdiscr}
\ee
where $r_X^{-1}(g) := (r_X(g))^{-1}$.

\begin{lemma}
\label{lemma:R_Xast-discr}
Let $X$ be a discrete vector field on a bicovariant group lattice. If $R_X$ is
invertible on $\A$, then  $R_{X \ast}$ maps discrete vector fields to
discrete vector fields.
\end{lemma}
{\bf Proof:} This is a simple consequence of (\ref{R_XastYdiscr}).
\hfill $\blacksquare$

\subsection{Basic vector fields}
\label{subsec:basicvf}
A discrete vector field $X$ which at every site has exactly one
outgoing and one incoming $X$-arrow will be called {\em basic}.
This means that for each $g \in G$ there is precisely one
$h \in S$ such that $X^h(g)=1$ and precisely one $h' \in S$
such that  $X^{h'}(g {h'}^{-1})=1$.
As a consequence, $\sum_{h \in S} X^h(g) = 1$ (and thus $X^e = 0$)
and $\sum_{h \in S} X^h(gh^{-1}) = 1$.

\begin{lemma}
\label{lemma:basvf-sr}
A discrete vector field $X$ is basic if and only if $I+X$
is invertible on $\A$ and $s_X$ has values in $S$.
\end{lemma}
{\bf Proof:} This is an immediate consequence of theorem~\ref{theor:I+X}
and the definition of basic vector fields.
\hfill $\blacksquare$
\bigskip

For a basic vector field, the bijection $\phi_X : G \rightarrow G$
given by $\phi_X(g) = g \, s_X(g)$ satisfies
\be
  \phi_X^\ast f = \sum_{h \in S} X^h \, R^\ast_h f  \qquad
  ( \forall f \in \A )  \; .
\ee
In particular, the vector fields $\ell_h$, $h \in S$, are basic
and we have $\phi_{\ell_{h}} = R_h$.
\vskip.2cm

A set of basic vector fields $\{ X_h | \, h \in S \}$ forms a
basis of $\X$ if for all $g \in G$ we have
$e^g \cdot \{ X_h | \, h \in S \} = e^g \cdot \{ \ell_h | \, h \in S \}$.
The parametrization by $S$ can be fixed by setting $s_{X_h}(e) = h$
where $s_{X_h}$ is the map $G \to S$ associated with $X_h$.
This yields indeed a unique parametrization
since at every site and hence also at $e$ there is exactly one
vector field $X_h$ for which $e^e \cdot X_h = e^e \cdot \ell_h$.
\vskip.2cm

The elements of the dual basis of 1-forms are determined by
$\langle X_h , \alpha^{h'} \rangle = \dl^{h'}_h$.
The coefficient matrices in
\be
  X_h = \sum_{h' \in S} X^{h'}{}_h \cdot \ell_{h'} \, , \qquad
 \alpha^h = \sum_{h' \in S} \alpha^h{}_{h'} \, \theta^{h'}
\ee
which mediate the change of basis are inverse to one another.
At each $g \in G$, these matrices act as permutations on $S$.
The dual basis 1-forms satisfy
$\alpha^h f = (\phi^\ast_{X_h} f) \, \alpha^h$.
Furthermore, $\sum_h \alpha^h = \theta$ and
\be
   \d f = \sum_{h \in S} (X_h f) \, \alpha^h \, .
\ee
\vskip.2cm

\begin{example}
\label{ex:bv-S3}
Let us consider ${\cal S}_3$ with $S = \{ (12), (13), (23) \}$
(see also example~\ref{ex:G=S3}).
Fig.~\ref{fig:bavefi} shows three vector fields which form a basis of
$\X$ and satisfy the above parametrization condition.
\begin{figure}
\begin{center}
\includegraphics[scale=.8]{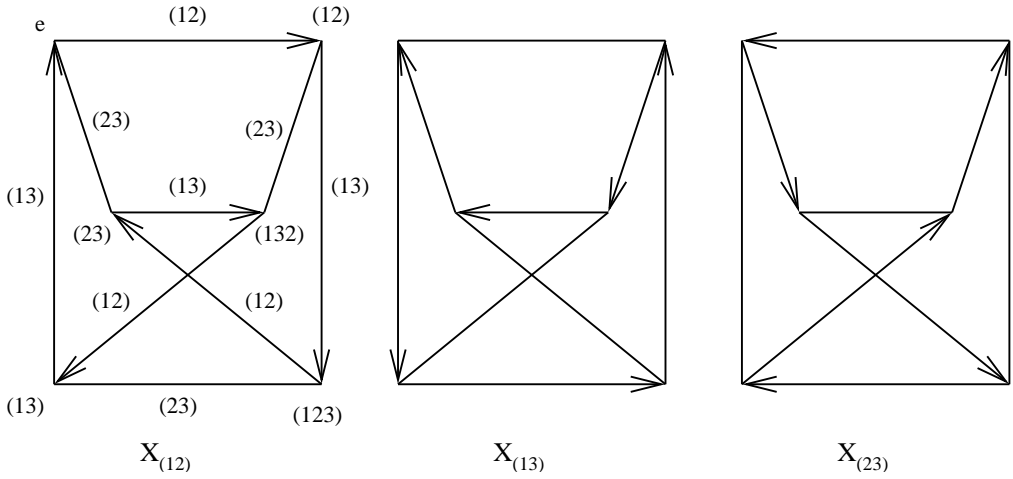}
\caption{A basis of basic vector fields on ${\cal S}_3$ with respect to
$S = \{ (12),(13),(23) \}$.}\label{fig:bavefi}
\end{center}
\end{figure}

For the basic vector field
\bez
  X = X_{(12)} = (e^{e}+e^{(123)}+e^{(132)}) \cdot \ell_{(12)}
  + (e^{(12)}+e^{(13)}+e^{(23)}) \cdot \ell_{(13)} \; .
\eez
we obtain $s_X(e)=s_X(123)=s_X(132)=(12)$ and
$s_X(12)=s_X(13)=s_X(23)=(13)$, and $\phi_X$ is
differentiable.\cite{bvS3}
Since $\theta^h = \sum_g e^g \, \d e^{gh}$, we have
$\phi_X^\ast (\theta^h) = \sum_g \phi_X^\ast (e^g) \, \d \phi_X^\ast
(e^{gh})$. In this way we obtain $\phi_X^\ast (\theta^{(12)}) =
\theta^{(13)}$. On the other hand, with
\bez
   R_X = (e^{e}+e^{(123)}+e^{(132)}) \, R^\ast_{(12)}
   + (e^{(12)}+e^{(13)}+e^{(23)}) \, R^\ast_{(13)} \; .
\eez
we find
\bez
  R_X (\theta^{(12)}) = (e^{e}+e^{(123)}+e^{(132)}) \, \theta^{(12)}
  + (e^{(12)}+e^{(13)}+e^{(23)}) \, \theta^{(23)}
\eez
which is obviously different from $\phi_X^\ast (\theta^{(12)})$.
Hence, in general we have $\phi_X^\ast \neq R_X$ on forms.
\end{example}

\begin{example}
\label{ex:bv-Z3Z3}
Let us choose $G = \mathbb{Z}_3 \times \mathbb{Z}_3$ with
$S = \{ (0,1), (1,0) \}$. Then
\bez
 X^{(0,1)} = e^{(0,0)} + e^{(1,0)} + e^{(0,1)} + e^{(2,1)} + e^{(1,2)} + e^{(2,2)}
   \, , \quad
 X^{(1,0)} = e^{(2,0)} + e^{(1,1)} + e^{(0,2)}
\eez
are the components of a basic vector field.
The corresponding map $\phi_X$ is {\em not} differentiable. Since
$(1,0)^{-1} (2,0) = (1,0) \in S$ (using a multiplicative notation
for the group operation), but
$s((1,0))^{-1} (1,0)^{-1} (2,0) \, s((2,0)) \not\in S_e$, this follows
using theorem~\ref{theor:phiXdiff}.
We can also apply (\ref{phi_diff}) directly: there is an arrow from
$(1,0)$ to $(2,0)$ in the group lattice, but $\phi_X((1,0)) = (1,1)$ is
not connected with $\phi_X((2,0)) = (0,0)$.
\end{example}

In the following we restrict our considerations to bicovariant group lattices
(so that $\ad(S)S \subset S$ and $\ad(S)^{-1}S \subset S$).

\begin{lemma}
\label{lemma:R_Xbas_inv}
If $X$ is a basic vector field, then $R_X$ is invertible on $\O$ with
\be
  R_X^{-1} = \sum_{h \in S} (R^\ast_{h^{-1}} X^h) \, R^\ast_{h^{-1}}
             \; .    \label{R_Xbas_inv}
\ee
Furthermore, for $Y \in \X$ we have
\be
  R_{X \ast} Y = \sum_{h \in S} (R^\ast_{h^{-1}} X^h) \, R_{h \ast} Y
      \; .    \label{bas:R_XastY}
\ee
\end{lemma}
{\bf Proof:} Since a basic vector field has an invertible flow and
$X^e = 0$, the first equation follows directly from lemma~\ref{lemma:R_X_inv}
and the second from lemma~\ref{lemma:R_XastY}.
\hfill $\blacksquare$
\bigskip

Two basic vector fields $X,Y$ form a \emph{biangle} if
$s_Y \cdot s_X = e$, which associates with each $g \in G$ a
group lattice biangle.\cite{s-not}
Three basic vector fields $X,Y,Z$ constitute a \emph{triangle} if
$s_Y \cdot s_X = s_Z$, which assigns to
each $g \in G$ a group lattice triangle. Furthermore,
four basic vector fields $X,Y,W,Z$ make up a \emph{quadrangle}
if $s_Y \cdot s_X = s_W \cdot s_Z \not\in S_e$.
This maps a group lattice quadrangle to each $g$. Below
we express these conditions more directly in terms of
the vector fields with the help of the next result.

\begin{lemma}
\label{lemma:R_XR_RXY}
For basic vector fields $X$ and $Y$ the following identity holds,
\be
   R_X \, R_{R_{X \ast} Y} = R^\ast_{s_Y \cdot s_X} \; .
        \label{R_XR_RXY}
\ee
\end{lemma}
{\bf Proof:}
\bez
     R_X \, R_{R_{X \ast} Y}
 &=& \sum_{h, h_1, h_2 \in S} X^{h_1} (R^\ast_{h_1 h^{-1}} X^h) \,
     (R^\ast_{h_1 h^{-1}} Y^{\ad(h)h_2}) \, R^\ast_{h_1 h_2}
  = \sum_{h_1, h_2 \in S} X^{h_1} Y^{\ad(h_1) h_2} R^\ast_{h_1h_2}  \\
 &=& \sum_{g \in S^2} \Big( \sum_{h_1,h_2 \in S} \dl^g_{h_1h_2} \, X^{h_1}
     Y^{\ad(h_1) h_2} \Big) \, R^\ast_g
  = \sum_{g \in S^2} \Big( \sum_{h_1,h_2 \in S} \dl^g_{h_2h_1} \, X^{h_1}
     Y^{h_2} \Big) \, R^\ast_g   \\
 &=& \sum_{g \in S^2} \dl^g_{s_Y \cdot s_X} \, R^\ast_g
  = R^\ast_{s_Y \cdot s_X}
\eez
using (\ref{R_XastY-expl}) and (\ref{XXdiscr-inv}).
\hfill $\blacksquare$
\bigskip

The above biangle condition is now seen to be equivalent to
\be
   R_X \, R_{R_{X \ast} Y} = I   \label{XY-biangle}
\ee
and the triangle condition can be expressed as
\be
   R_X \, R_{R_{X \ast} Y} = R_Z \, .
\ee
Furthermore, the above quadrangle condition takes the form
\be
  R_X \, R_{R_{X \ast} Y} = R_Z \, R_{R_{Z \ast} W} \neq R_{X'}
\ee
for all discrete vector fields $X'$.

\subsection{Lie derivative with respect to a discrete vector field}
The notion of the Lie derivative is easily taken over from continuum
differential geometry to the discrete framework of
group lattices. Let $X$ be a discrete vector field. On functions,
the Lie derivative with respect to $X$ is given by
\be
  \L_X f = \phi^\ast_X f - f = (I+X)f - f = Xf  \, .
\ee
If $\phi_X$ is differentiable, we can extend the Lie derivative
to forms $\omega \in \O$ via
\be
      \L_X \omega = \phi_X^\ast \omega - \omega
      \label{Lie_X_omega}
\ee
so that, in particular,
\be
  \L_{\ell_h} \omega = R_h^\ast \omega - \omega \, , \qquad
  \L_{\ell_h} \theta^{h'} = \theta^{\ad(h)h'} - \theta^{h'} \, .
\ee
 For $\psi, \omega \in \O$, we also have
\be
     \L_X (\psi \, \omega)
  = \phi^\ast_X (\psi \, \omega) - \psi \, \omega
  = (\L_X \psi) \, \omega + \psi \, \L_X \omega
     + (\L_X \psi) \, \L_X \omega \, .
\ee
\vskip.1cm

Assuming $\phi_X$ to be differentiable and invertible, the
Lie derivative acts on vector fields as follows,
\be
   \L_X Y = Y - \phi_{X\ast} Y = \phi^\ast_X{}^{-1} \circ [X,Y]
\ee
using (\ref{phi_astX}).
In particular, with $I + \ell_h = R^\ast_h$ we obtain
\be
    \L_{\ell_h} \ell_{h'}
  = R_{h^{-1}}^\ast \circ [ R_h^\ast , R_{h'}^\ast ]
  = R_{h'}^\ast - R_{\ad(h^{-1})h'}^\ast \; .
\ee
Since $\ad(h^{-1})h' \in S$ for differentiable $R_h$,
this can be written as
\be
   \L_{\ell_h} \ell_{h'} = \ell_{h'} - \ell_{\ad(h^{-1})h'}
   \label{liell}
\ee
and also in the form
\be
    \L_{\ell_h} \ell_{h'}
  = \ell_h \, \ell_{\ad(h^{-1})h'} - \ell_{h'} \, \ell_h
\ee
which involves a generalization of the ordinary commutator
of vector fields.

\subsection{Inner product of discrete vector fields and forms}
In this subsection we extend the inner product (or contraction)
$\langle X , \alpha \rangle$ of vector fields and 1-forms to
forms of higher grade. More precisely, we restrict our
considerations to {\em discrete} vector fields $X$ with a
{\em differentiable} flow, i.e., the associated map $\phi_X$ is
assumed to be differentiable.
\vskip.1cm

For all $f \in \A$ and $\alpha \in \O^1$ we require
\be
   X \lc f = 0 \, , \qquad
   X \lc \alpha  = \langle X , \alpha \rangle \; .
        \label{ip_X_f}
\ee
Furthermore, for a discrete vector field $X$ with differentiable
map $\phi_X$, we demand
\be
   X \lc ( \omega \, \omega' )
 = (X \lc \omega) \, \phi_X^\ast \omega' + (-1)^r \, \omega \, (X \lc \omega')
     \label{ip_X_Leibniz}
\ee
for all $\omega \in \O^r$ and $\omega' \in \O$.\cite{innprod}
In particular, using (\ref{X_df}) and the Leibniz rule for $\d$, we obtain
\be
     X ( f \, f')
   = X \lc \d (f \, f')
   = (X f) \, \phi_X^\ast f' + f \, X f'
\ee
which is a reformulation of (\ref{discr_vf}).
\vskip.1cm

It is easily verified that (\ref{ip_X_Leibniz}) is compatible with
the $\A$-bimodule structure of $\O$, i.e.
$X \lc [ (\omega f) \, \omega' ] = X \lc [ \omega \, (f \omega') ]$.
The consistency with the commutation relations (\ref{theta_f})
follows from
\be
   X \lc ( \theta^h \, f )
 &=& ( X \lc \theta^h ) \, \phi_X^\ast f
  = X^h \sum_{h'} X^{h'} \, R_{h'}^\ast f
  = X^h \, R_h^\ast f
  = (R_h^\ast f) \, (X \lc \theta^h)  \nonumber \\
 &=& X \lc [ (R_h^\ast f) \, \theta^h ]
\ee
which holds for a discrete vector field $X$.
\vskip.1cm

The definition (\ref{ip_X_Leibniz}) is also consistent with
the 2-form relations. Let $(g,g') \in {\cal I}$, so that
$0 = e^g \, \d e^{g'} = - (\d e^g) \, e^{g'}$.
The corresponding 2-form relation is
$\d e^g \, \d e^{g'} = 0$. Applying $X \lc$ to the left hand side,
we obtain
\be
     X \lc (\d e^g \, \d e^{g'})
 &=& (X \lc \d e^g) \, \phi_X^\ast (\d e^{g'})
     - \d e^g \, X \lc \d e^{g'}
  = (X e^g) \, \d( \phi_X^\ast e^{g'}) - \d e^g \, (X e^{g'})
              \nonumber \\
 &=& (\phi_X^\ast e^g - e^g) \, \d( \phi_X^\ast e^{g'})
     - \d e^g \, ( \phi_X^\ast e^{g'} - e^{g'}) \nonumber \\
 &=& \phi_X^\ast (e^g \, \d e^{g'}) - e^g \, \d( \phi_X^\ast e^{g'})
     - \d e^g \, \phi_X^\ast e^{g'} - (\d e^g) \, e^{g'} \nonumber \\
 &=& - \d( e^g \, \phi_X^\ast e^{g'} ) \; .
\ee
But the last expression vanishes since the function
$e^g \, \phi_X^\ast e^{g'}$ vanishes identically. Indeed, it obviously
vanishes at elements of $G$ different from $g$. Evaluated at $g$,
it yields
$(\phi_X^\ast e^{g'})(g) = e^{g'}(\phi_X(g)) = e^{g'}(g \, s(g))$
which vanishes since $(g,g') \in {\cal I}$.

\vskip.1cm
\noindent
{\em Remark.} Let $h_1 h_2 = h_2 h_3 = \cdots = h_r h_1$ be a cycle
of a bicovariant group lattice. Then
\bez
   \ell_h \lc (\theta^{h_1} \theta^{h_2}
   + \theta^{h_2} \theta^{h_3} + \ldots)
 = \dl^{h_1}_h R^\ast_h \theta^{h_2}
 - \dl^{h_2}_h \theta^{h_1}
 + \dl^{h_2}_h R^\ast_h \theta^{h_3}
 - \dl^{h_3}_h \theta^{h_2} + \ldots
\eez
where the second and the third term on the right-hand side cancel
since $\ad(h_2) h_3 = h_1$, and the same happens with the remaining terms.
In particular, the first term cancels the last one.
Since the 2-form relations are sums of cycles, this means that $\ell_h \lc$
applied to a 2-form relation automatically vanishes. In fact, we have
the stronger result that $\ell_h$-contractions with any cycle vanish (which
perfectly matches the Woronowicz wedge product).
A particular consequence is that the condition $\psi = 0$ for a 2-form
$\psi = \psi_{h_1,h_2} \, \theta^{h_1} \theta^{h_2}$ is stronger than
$\ell_h \lc \psi = 0$ for all $h \in S$.
For example, the vanishing of $\ell_{h_1} \lc \ell_{h_2} \lc \psi
 = \psi_{h_2,\ad(h_2^{-1})h_1} - \psi_{h_1,h_2}$
obviously does not imply vanishing $\psi$.
\hfill $\blacksquare$

\begin{lemma}
\label{lemma:X_Delta}
If $X$ is a basic vector field with differentiable
flow, then
\be
   X \lc \Delta(\omega) + \Delta( X \lc \omega) = 0
   \qquad ( \forall \omega \in \O ) \; .   \label{X_Delta}
\ee
\end{lemma}
{\bf Proof:} For functions the identity is trivially satisfied. Let us
prove it first for 1-forms. A basic vector field satisfies
$X \lc \theta = \mathbf{1}$ and the flow map $\phi_X$ is a bijection.
Since $\phi_X$ is assumed to be differentiable,
we also have $\phi_X^\ast \theta = \theta$ according to
(\ref{phiast_theta}). As a consequence, we find
\bez
     X \lc \theta^2
  = (X \lc \theta) \, \phi_X^\ast \theta
    - \theta \, (X \lc \theta)
  = 0 \; .
\eez
Using
\bez
  \Delta ( [\theta, f] ) = [ \Delta (\theta) , f ]
                         = [\theta^2 - \Delta^e, f]
                         = [\theta^2, f]
\eez
for $f \in \A$, we thus obtain
\bez
  X \lc \Delta (\d f) + \Delta ( X \lc \d f ) = X \lc \Delta (\d f) = 0 \; .
\eez
Since every 1-form $\alpha$ is a sum of terms like $f' \, \d f$
with $f,f' \in \A$, the last identity extends to
\bez
  X \lc \Delta(\alpha) + \Delta(X \lc \alpha) = X \lc \Delta(\alpha) = 0 \, .
\eez
Let us now assume that our assertion holds for differential forms of grade
$< k$. Then we find
\bez
  & & X \lc \Delta(\psi \, \omega) + \Delta(X \lc (\psi \, \omega) ) \\
  &=& X \lc \Big( \Delta(\psi) \, \omega + (-1)^r \, \psi \, \Delta(\omega) \Big)
     + \Delta \Big((X \lc \psi) \, \phi_X^\ast \omega
     + (-1)^r \, \psi X \lc \omega \Big)   \\
  &=& X \lc \Delta(\psi) \, \phi_X^\ast \omega
     - (-1)^r \, \Delta(\psi) \, X \lc \omega
     + (-1)^r \, (X \lc \psi) \, \phi_X^\ast \Delta(\omega)
     + \psi \, X \lc \Delta(\omega)  \\
  & &+ \Delta(X \lc \psi) \, \phi_X^\ast \omega
     - (-1)^r \, (X \lc \psi) \, \Delta(\phi_X^\ast \omega)
     + (-1)^r \, \Delta(\psi) \, X \lc \omega + \psi \, \Delta(X \lc \omega) \\
  &=& 0
\eez
for $\psi \in \O^r$, $r< k$, and $\omega \in \O^{<k}$, using
$\phi_X^\ast \circ \Delta = \Delta \circ \phi_X^\ast$
(see (\ref{Delta_phiast})). By induction on the
grade of the argument $\omega$, the formula (\ref{X_Delta}) is proven.
\hfill $\blacksquare$
\bigskip

\begin{theorem}
\label{theor:Lie-Cartan}
For a basic vector field $X$ with differentiable flow the
following (Lie-Cartan) identity holds,
\be
   \L_X \omega = X \lc \d \omega + \d ( X \lc \omega )
   \qquad ( \forall \omega \in \O ) \; .
              \label{Lie_Cartan}
\ee
\end{theorem}

\noindent
{\bf Proof:} With the help of (\ref{domega_Delta}) and (\ref{ip_X_Leibniz}),
(\ref{X_Delta}) can be reformulated as follows,
\bez
  0 &=& X \lc \Delta(\omega) + \Delta(X \lc \omega)  \\
    &=& X \lc ( [\theta, \omega] - \d \omega ) + [\theta, X \lc \omega]
        - \d (X \lc \omega)  \\
    &=& (X \lc \theta) \, \phi_X^\ast \omega
        - (-1)^r \, (X \lc \omega) \, \phi_X^\ast \theta
        - \omega \, X \lc \theta - X \lc \d \omega
        + (-1)^r \, (X \lc \omega) \, \theta - \d (X \lc \omega) \\
    &=& \phi_X^\ast \omega - \omega - X \lc \d \omega - \d (X \lc \omega)
\eez
for $\omega \in \O^r$, using in the last step $X \lc \theta = \mathbf{1}$
and $\phi_X^\ast \theta = \theta$ which hold for a basic vector field with
differentiable flow. Now (\ref{Lie_Cartan}) is obtained recalling
the definition (\ref{Lie_X_omega}).
\hfill $\blacksquare$

\begin{lemma}
\label{lemma:phiX_omega}
If $\phi : G \rightarrow G$ is an invertible differentiable map
and $X$ a discrete vector field with differentiable flow, then
\be
  \phi^\ast (X \lc \omega) = (\phi^{-1}_\ast X) \lc \phi^\ast \omega
  \qquad ( \forall \omega \in \O ) \, .
\ee
\end{lemma}
{\bf Proof:} For a 1-form $\alpha$ the formula follows from
(\ref{phi_astX_def}) (even more generally for an arbitrary vector field $X$).
Furthermore, we have
\bez
     \phi^\ast [X \lc (\psi \, \omega)]
 &=& \phi^\ast [(X \lc \psi) \, \phi^\ast_X \omega + (-1)^r \, \psi \, X \lc \omega] \\
 &=& \phi^\ast (X \lc \psi) \, \phi^\ast \phi^\ast_X \omega
    + (-1)^r \, (\phi^\ast \psi) \, \phi^\ast (X \lc \omega) \, .
\eez
Let us assume that the identity holds for grades lower than $k$.
For $\psi \in \O^r$, $r<k$, and $\omega$ of grade lower than $k$,
we then obtain
\bez
     \phi^\ast [X \lc (\psi \, \omega)]
 &=& [(\phi^{-1}_\ast X) \lc \phi^\ast \psi] \, \phi^\ast \phi^\ast_X \phi^{\ast-1}
     (\phi^\ast \omega)
     + (-1)^r \, (\phi^\ast \psi) \, (\phi^{-1}_\ast X) \lc \phi^\ast \omega  \\
 &=& (\phi^{-1}_\ast X) \lc \phi^\ast (\psi \, \omega)
\eez
since
\bez
    \phi^\ast \phi^\ast_X \phi^{\ast-1} = \phi^\ast (I+X) \phi^{\ast-1}
  = I + \phi^{-1}_\ast X = \phi_{\phi^{-1}_\ast X}^\ast
\eez
using (\ref{phi_astX}). Now the identity follows by induction.
\hfill $\blacksquare$

\begin{lemma}
\label{lemma:XX_omega}
A discrete vector field $X$ with differentiable flow
satisfies
\be
    X \lc X \lc \omega = 0
      \qquad ( \forall \omega \in \O ) \; .
\ee
\end{lemma}
{\bf Proof:} Again, we use induction with respect to the grade
of $\omega$.
We have $X \lc X \lc \alpha = 0$ trivially for $\alpha \in \O^1$.
Next we calculate
\bez
     X \lc X \lc (\psi \, \omega)
 &=& X \lc [(X \lc \psi) \, \phi^\ast_X \omega
     + (-1)^r \, \psi X \lc \omega]  \\
 &=& (X \lc X \lc \psi) \, \phi^{\ast 2}_X \omega
     - (-1)^r \, (X \lc \psi) \, X \lc \phi^\ast_X \omega \\
 & & + (-1)^r \, (X \lc \psi) \, \phi^\ast_X (X \lc \omega)
     + \psi (X \lc X \lc \omega)
\eez
with the help of lemma~\ref{lemma:phiX_omega} and
\bez
   \phi^{-1}_{X \ast} X = \phi_X^\ast \, X \, (\phi_X^{-1})^\ast
   = \phi_X^\ast \, (\phi_X^\ast - I) \, (\phi_X^{-1})^\ast
   = \phi_X^\ast - I = X \; .
\eez
This implies that if the assertion
holds for $\omega$ of grade $<r$, then it also holds for grade $r$.
\hfill $\blacksquare$

\section{Connections and parallel transports}
\label{sec:connections}
\setcounter{equation}{0}
A {\em connection} on a left $\A$-module $\E$ is a linear map
$\nabla : \E \rightarrow \O^1 \oA \E$ such that
\be
   \nabla (f \, E) = \d f \oA E + f \, \nabla(E)  \qquad (\forall E \in \E ) \, .
   \label{nafE}
\ee
If $(\O,\d)$ is the differential calculus of a group lattice, we have the
following result.

\begin{lemma}
\label{lemma:connE}
Every connection on $\E$ is of the form
\be
  \nabla(E) = \theta \oA E - \V(E)  \qquad (\forall E \in \E )
              \label{na_E}
\ee
where $\V : \E \rightarrow \O^1 \oA \E$ satisfies
\be
    \V( f \, E ) = f \, \V(E) \, . \label{V_left_tensor_prop}
\ee
Conversely, every linear map $\V$ with this property defines a
connection via the above formula.
\end{lemma}
{\bf Proof:} This is easily verfied using (\ref{df_theta}).
\hfill $\blacksquare$
\bigskip

Writing
\be
    \V = \sum_{h \in S} \theta^h \oA \V_{\ell_h}
         \label{V_Vellh}
\ee
with {\em parallel transport operators} $\V_{\ell_h}$ in the
$\ell_h$ direction, (\ref{V_left_tensor_prop}) leads to
\be
  \V_{\ell_h}( f \, E) = (R^\ast_{h^{-1}} f) \, \V_{\ell_h}(E)
        \label{Vellh_fE}
\ee
using (\ref{theta_f}). In particular,
\be
   \V_{\ell_h}(e^g \, E) = e^{gh} \, \V_{\ell_h}(E)
\ee
which shows that we have a transport in the forward direction.
We generalize it to a transport along an arbitrary
vector field $X$ by
\be
  \V_X = \sum_{h \in S} (R^\ast_{h^{-1}} X^h) \, \V_{\ell_h}
         \, .         \label{V_X-E}
\ee

\begin{lemma}
\label{lemma:V_basX}
For a basic vector field $X$,
\be
   \V_X (f \, E) = (R_X^{-1} f) \, \V_X E \; .
       \label{V_basicX(fE)}
\ee
\end{lemma}
{\bf Proof:} Using (\ref{V_X-E}), (\ref{Vellh_fE}), (\ref{XXdiscr-inv}) and
(\ref{R_Xbas_inv}) we obtain
\bez
     \V_X (f \, E)
 &=& \sum_{h \in S} (R^\ast_{h^{-1}} X^h) \, (R^\ast_{h^{-1}} f)
     \, \V_{\ell_h} E  \\
 &=& \sum_{h,h' \in S} (R^\ast_{h^{-1}} X^h) \, (R^\ast_{h^{-1}} f) \,
     (R^\ast_{{h'}^{-1}} X^{h'}) \, \V_{\ell_{h'}} E \\
 &=& (R_X^{-1} f) \, \V_X E \; .
\eez
\hfill $\blacksquare$
\bigskip

A connection can be extended to a map $\na : \O \oA \E \to \O \oA \E$
via
\be
   \na(\omega \oA E)
 = \d \omega \oA E + (-1)^r \omega \, \na E  \qquad
   \forall \omega \in \O^r, \, E \in \E  \, .
      \label{nabla_ext}
\ee
The {\em curvature} of the connection $\na$ is the left
$\A$-module homomorphism $\R : \E \to \O^2 \oA \E$ defined by
\be
   \R(E) = - \na^2 E  \; .
\ee
More generally, $\R = - \nabla^2$ is defined as a map
$\O \oA \E \to \O \oA \E$. It has the property
\be
    \R(\omega \oA E) = \omega \, \R(E)
\ee
and satisfies the \emph{second Bianchi identity}
\be
   (\na \R)(E) := \na (\R(E)) - \R (\na E) = 0 \; .
      \label{2ndBianchi}
\ee

\subsection{Gauge theory}
Let $\E$ be a right $\A$-module. Then $\E \, e^g$, for fixed $g \in G$,
is a complex vector space. Let $E_i(g)$, $i=1,\ldots,m(g)$, be a basis
of this vector space. In general, its dimension varies with $g$.
In the following we assume, for simplicity, that $m(g)$ is independent
of $g$ and finite.\cite{bundles}
Choosing an order $E_1(g),\ldots,E_m(g)$ for all $g \in G$,
we obtain a right $\A$-module basis of $\E$ by setting
$E_i := \sum_{g \in G} E_i(g)$. Then $\E$ is a free right $\A$-module.
\vskip.1cm

An element $\Psi \in \E \oA \O^r$ can be written as
$\Psi = E_i \oA \psi^i$ (using the summation convention) with an
$r$-form field $\psi : G \to (\O^r)^m$ transforming
according to $\psi \mapsto \psi' = \gamma \psi$ under the action of a
gauge group $\Gamma$, corresponding to changes of the basis of $\E$.
A {\em right} $\A$-module connection $\na$ has to satisfy
$\na(E \oA \omega) = \na(E) \, \omega + E \oA \d \omega$
for all $\omega \in \O$, so that
\be
  \na \Psi = \na(E_j) \, \psi^j + E_i \oA \d \psi^i
           = E_i \oA ( \d \psi^i + A^i{}_j \, \psi^j)
           = E_i \oA \D \psi^i  \; .
\ee
Here $A$ is a gauge potential 1-form and
\be
  \D \psi := \d \psi + A \, \psi
\ee
the \emph{exterior covariant derivative} of $\psi$ with the
transformation law $(\D \psi)' = \gamma \, \D \psi$.
\vskip.2cm

Similarly, an $r$-form field $\varphi$ transforming according to
$\varphi \mapsto \varphi' = \varphi \gamma^{-1}$ under the action
of the gauge group corresponds to an element of a {\em left}
$\A$-module. Then
\be
    \D \varphi := \d \varphi - (-1)^r \varphi A
\ee
defines a covariant exterior derivative, i.e.,
$(\D \varphi)' = (\D \varphi) \, \gamma^{-1}$. Furthermore, we have
\be
    (\D \varphi) \, \psi + (-1)^r \, \varphi \, \D \psi
  = \d (\varphi \, \psi) \; .
\ee
Introducing
\be
   W := \theta + A = \sum_{h \in S} W_h \, \theta^h
\ee
which obeys the transformation law
\be
  W \to W' = \gamma \, W \, \gamma^{-1} \, , \qquad
      W'_h = \gamma \, W_h \, (R^\ast_h \gamma^{-1})
\ee
under a gauge transformation, and using (\ref{domega_Delta}),
we obtain
\be
    \D \psi &=& W \psi - (-1)^r \psi \theta - \Delta(\psi)  \\
 \D \varphi &=& \theta \varphi - (-1)^r \varphi W - \Delta(\varphi) \; .
\ee
\vskip.2cm

 From $\na^2 \Psi = E_i \oA \D^2 \psi =: E_i \oA F^i{}_j \, \psi^j$
originates the curvature 2-form
\be
  F = \d A + A^2 = W^2 - \Delta(W) - \Delta^e
    = \sum_{h,h' \in S} F_{h,h'} \, \theta^h \theta^{h'}
\ee
which satisfies the Bianchi identity
\be
   0 = \D F := \d F + [A, F] = [W,F] - \Delta(F) \; .
\ee
The biangle, triangle and quadrangle parts of the curvature 2-form are,
respectively, given by
\be
 F_{(e) \, h,h'} = W_h \, R^\ast_h W_{h'}-I
     \quad && \mbox{for a biangle $hh'=e$}  \\
 F_{(h_0) \, h,h'} = W_h \, R^\ast_h W_{h'} - W_{h_0}
     \quad && \mbox{for a triangle $hh'=h_0 \in S_{(1)}$} \\
 F_{(g) \,h,h';\hat{h},\hat{h}'} =   W_h \, R^\ast_h W_{h'}
   - W_{\hat{h}} \, R^\ast_{\hat{h}}W_{\hat{h}'}
     \quad &&
  \mbox{for a quadrangle $hh'=\hat{h}\hat{h}'=g \in S_{(2)}$}
  \, . \qquad
\ee
\vskip.2cm

For 0-form fields $\psi$ and $\phi$ we write
\be
   \D \psi = \sum_{h \in S} \na_{\ell_h} \psi \; \theta^h \, , \qquad
   \D \varphi = \sum_{h \in S} (\na_{\ell_h} \varphi) \, W_h \, \theta^h
   \, .
\ee
Then
\be
  \na_{\ell_h} \psi = W_h \, R^\ast_h \psi - \psi \, .
\ee
If the group $\Gamma$ is unitary and if $W_h^{-1} = W_h^\dagger$,
(where ${}^\dagger$ denotes hermitian conjugation) then
$\psi^\dagger \mapsto \psi^\dagger \, \gamma^{-1}$ and we obtain
\be
    \na_{\ell_h} \psi^\dagger
  = (R^\ast_h \psi^\dagger) \, W_h^{-1} - \psi^\dagger
  = (\na_{\ell_h} \psi)^\dagger \; .
\ee
An example of a Lagrangian for the 0-form field $\psi$ is
\be
     {\cal L}_\psi
  = \sum_{h \in S} {1 \over 2} \na_{\ell_h} \psi^\dagger \,
    \na_{\ell_h} \psi
  = \sum_{h \in S} {1 \over 2} \Big(R^\ast_h (\psi^\dagger \psi)
     + \psi^\dagger \psi - \psi^\dagger W_h \, R^\ast_h \psi
     - (R^\ast_h \psi^\dagger) W^\dagger_h \psi\Big)
\ee
with corresponding action
\be
     \mathbb{S}_\psi
  = \sum_{g \in G} {\cal L}_\psi (g)
  = \sum_{g \in G} \sum_{h \in S} {1 \over 2} \Big( 2 \psi^\dagger \psi
     - \psi^\dagger W_h \, R^\ast_h \psi
     - \psi^\dagger (R^\ast_{h^{-1}} W^\dagger_h \psi) \Big)(g) \, .
\ee
\vskip.2cm

In order to build a Lagrangian from $r$-form fields, $r>0$, we need an
inner product of $r$-forms. It should satisfy
\be
   (f \, \omega, f' \, \omega') = f^\dagger \, f' \, (\omega, \omega')
\ee
(where $f^\dagger$ is the complex conjugate of the function $f$).
A natural choice of inner product of 1-forms is then determined by
\be
   (\theta^h, \theta^{h'}) = \dl^{h,h'} \; .
\ee
As a consequence, the above Lagrangian for a 0-form field $\psi$
can be written as follows,
\be
    {\cal L}_\psi = {1 \over 2} \, (\D\psi^\dagger , \D\psi) \; .
\ee
\vskip.1cm

For a biangle or triangle $h_1h_2 \in S_e$, we set
\be
    (\theta^{h_1} \theta^{h_2} , \theta^h \theta^{h'})
  = \dl^{h_1,h} \, \dl^{h_2,h'} \; .
\ee
For a quadrangle $h_1h_2=g \not\in S_e$ and a 2-form $\psi$ we
define
\be
  (\theta^{h_1} \theta^{h_2}, \psi) = \psi_{(g) h_1,h_2}
\ee
where $\psi_{(g) h_1,h_2}$ are the quadrangle components of
$\psi$ as defined in (\ref{psi_quadr_comp}). In particular,
\be
    (\theta^{h_1} \theta^{h_2}, \theta^h \, \theta^{h'})
  = |g| \, \dl^{h_1,h} \, \dl^{h_2,h'} - \dl^{hh'}_g
    \quad \mbox{if } h_1h_2 = g \in S_{(2)}  \; .
\ee
As a consequence of these definitions, biangle, triangle and
quadrangle 2-forms are orthogonal to each other.
The Yang-Mills Lagrangian for the gauge potential $A$ then
takes the form
\be
  {\cal L}_{\rm YM} := {1 \over 2 m} \, {\rm tr} \Big(
  (p_{(e)} F , p_{(e)}F) + \sum_{h \in S_{(1)}} (p_{(h)}F , p_{(h)}F)
  + \sum_{g \in S_{(2)}} {1 \over |g|}(p_{(g)}F , p_{(g)}F) \Big)
\ee
and the corresponding action is
$\mathbb{S}_{\rm YM} = \sum_{g' \in G} {\cal L}_{\rm YM}(g')$.
 From biangles, triangles and quadrangles, respectively,
the following contributions arise:
\be
     {\rm tr}(p_{(e)}F,p_{(e)}F)
 &=& \sum_{h,h' \in S} \dl^e_{hh'} \, {\rm tr} \Big( 2I - W_h \, (R^\ast_h W_{h'})
     - (R^\ast_h W_{h'}^\dagger ) \, W_h^\dagger \Big) \, , \\
     {\rm tr}(p_{(h_0)}F,p_{(h_0)}F)
 &=& \sum_{h,h' \in S} \dl^{h_0}_{hh'} \, {\rm tr} \Big( 2I
     - W_{h_0}^\dagger W_h \, (R^\ast_h  W_{h'})
     - (R^\ast_h W_{h'}^\dagger) \, W_h^\dagger W_{h_0} \Big) \, ,  \\
     {\rm tr}(p_{(g)}F,p_{(g)}F)
 &=& {\rm tr} \Big( 3|g|I - \sum_{h_1,h_2,h_3,h_4 \in S} \dl^g_{h_1h_2} \,
      \dl^g_{h_3h_4} \, (R^\ast_{h_1} W_{h_2}^\dagger) \,
      W_{h_1}^\dagger W_{h_3} \, (R^\ast_{h_3} W_{h_4}) \Big) . \qquad
\ee
These expressions are indeed gauge invariant and thus also
${\cal L}_{\rm YM}$. The latter generalizes the Lagrangian of
lattice gauge theory to arbitrary group lattices. It involves
parallel transports $U_P$ around the special plaquettes $P$ given
by biangles, triangles and quadrangles. Lattice gauge theory models
on group lattices $(G,S)$ with $S=S^{-1}$ have previously
been considered in Ref. \cite{Lech+Samu95} with an action of the
form $\sum_{P \in {\cal P}} {\rm tr} [U_P + U_P^{-1}]$ where the
sum is over some choice of set $\cal P$ of plaquettes (not
restricted to biangles, triangles and quadrangles). In contrast,
we have used the natural differential geometry of the group lattice
in order to determine a direct analogue of the Yang-Mills action.

\section{Linear connections}
\label{sec:lincon}
\setcounter{equation}{0}
A connection on $\O^1$, regarded as a left $\A$-module, is called
a {\em linear connection}. We introduce matrices
$V_h = (V^{h''}{}_{h,h'})$ with entries in $\A$ via
\be
   \V_{\ell_{h'}} (\theta^h)
 = \sum_{h'' \in S} (R^\ast_{{h'}^{-1}} V^h{}_{h',h''}) \, \theta^{h''}
        \label{Vellh_thetah}
\ee
so that
\be
  \nabla \theta^h = \theta \oA \theta^h - \sum_{h' \in S} V^h{}_{h'} \oA \theta^{h'}
     \label{nablathetah}
\ee
with
\be
    V^h{}_{h'} := \sum_{h'' \in S} V^h{}_{h'',h'} \, \theta^{h''} \; .
\ee
From the definition of the curvature we obtain
\be
   \R(\theta^h) = - \Delta^e \oA \theta^h
 - \sum_{h' \in S} \Delta(\theta^{h'}) \oA \V_{\ell_{h'}}(\theta^h)
 + \sum_{h',h'' \in S} \theta^{h'} \theta^{h''}
   \oA \V_{\ell_{h''}} \V_{\ell_{h'}}(\theta^h)
          \label{Rthetah}
\ee
where we used (\ref{dthetah}), (\ref{dtheta}), (\ref{na_E})
and (\ref{V_Vellh}).
\vskip.2cm

The {\em torsion} of a linear connection is the left $\A$-module
homomorphism $\Theta : \O^1 \to \O^2$ defined by
\be
 \Theta(\alpha) = \d \alpha - \pi \circ \na \alpha  \qquad \qquad
 \forall \alpha \in \O^1
\ee
where $\pi$ is the canonical projection $\O^1 \oA \O^1 \rightarrow \O^2$.
Then
\be
   \Theta^h &:=& \Theta(\theta^h)
 = \theta^h \, \theta - \Delta(\theta^h) + \pi \V(\theta^h)
 = \theta^h \, \theta - \Delta(\theta^h)
   + \sum_{h' \in S} \theta^{h'} \V_{\ell_{h'}}(\theta^h) \nonumber \\
&=& \sum_{h_1,h_2 \in S} (\dl^h_{h_1} - \dl^h_{h_1h_2}
   + V^h{}_{h_1,h_2}) \, \theta^{h_1} \theta^{h_2}
             \label{Thetah}
\ee
using (\ref{dthetah}), (\ref{Delta}), (\ref{na_E}), (\ref{V_Vellh})
and (\ref{Vellh_thetah}).
The torsion extends to a map $\Theta : \O \oA \O^1 \to \O$ via
\be
   \Theta = \d \circ \pi - \pi \circ \na   \label{torsion2}
\ee
where $\pi$ now denotes more generally the canonical projection
$\O \oA \O^1 \to \O$. It has the property
\be
  \Theta (\omega \oA \alpha) = (-1)^r \, \omega \, \Theta(\alpha)
\ee
for all $\alpha \in \O^1$ and $\omega \in \O^r$. From (\ref{torsion2})
we obtain the \emph{first Bianchi identity}
\be
   \d \circ \Theta + \Theta \circ \na = \pi \circ \R
\ee
and thus
\be
   \d \Theta^h - \theta \, \Theta^h
   + \sum_{h' \in S} V^h{}_{h'} \, \Theta^{h'}
 = \pi \, \R(\theta^h)
\ee
or, using (\ref{domega_Delta}),
\be
    \Theta^h \, \theta + \Delta(\Theta^h)
   - \sum_{h' \in S} V^h{}_{h'} \, \Theta^{h'}
 = - \pi \, \R(\theta^h)  \; .  \label{1stBianchi^h}
\ee
Writing
\be
  \R(\theta^h) = \sum_{h' \in S} \R^h{}_{h'} \oA \theta^{h'}
      \label{Rhh'}
\ee
with $\R^h{}_{h'} \in \O^2$, we find
\be
   \nabla ( \R(\theta^h) )
 = \sum_{h' \in S} \Big( \theta \, \R^h{}_{h'} - \Delta( R^h{}_{h'} )
   - \sum_{h'' \in S} \R^h{}_{h''} \, V^{h''}{}_{h'} \Big) \oA \theta^{h'}
\ee
using (\ref{nabla_ext}), (\ref{Delta(omega)}) and (\ref{nablathetah}).
Furthermore,
\be
   \R( \nabla \theta^h ) = \sum_{h' \in S} \Big( \theta \, \R^h{}_{h'}
   - \sum_{h'' \in S} V^h{}_{h''} \, \R^{h''}{}_{h'} \Big) \oA \theta^{h'}
\ee
so that the second Bianchi identity (\ref{2ndBianchi}) takes the form
\be
   \Delta(\R^h{}_{h'}) = \sum_{h'' \in S} ( V^h{}_{h''} \, \R^{h''}{}_{h'}
                          - \R^h{}_{h''} \, V^{h''}{}_{h'} ) \; .
\ee

\subsection{A transport of vector fields}
\label{subsec:trvf}
Let $\na : \O^1 \rightarrow \O^1 \oA \O^1$ be a linear connection
with parallel transport operator $\V_X$. Via
\be
   \langle \tilde{\V}_{\ell_h} Y, \alpha \rangle
 = R^\ast_h \langle Y, \V_{\ell_h} \alpha \rangle \, .
            \label{tV}
\ee
a dual of $\V_{\ell_h}$ is defined which acts on vector fields.
 From this definition we obtain
\be
   \tilde{\V}_{\ell_h} (f \cdot Y)
 = (R^\ast_h f) \cdot \tilde{\V}_{\ell_h} Y \; .
       \label{tV(fY)}
\ee
In particular,
\be
    \tilde{\V}_{\ell_h} (e^g \cdot Y)
  = e^{gh^{-1}} \cdot \tilde{\V}_{\ell_h} Y
\ee
which shows that the transport acts in the backward direction
$gh^{-1} \leftarrow g$. Furthermore, (\ref{Vellh_thetah}) leads to
\be
   \tilde{\V}_{\ell_h} \ell_{h'}
 = \sum_{h'' \in S} V^{h''}{}_{h,h'} \cdot \ell_{h''} \, .
       \label{tVell-ell}
\ee
Defining
\be
   \tilde{\V}_X := \sum_{h \in S} X^h \, \tilde{\V}_{\ell_h}
       \qquad ( \, \mbox{on } \X \, )   \label{tV_X}
\ee
(\ref{tV}) and (\ref{tV(fY)}) generalize, respectively, to
\be
   \langle \tilde{\V}_X Y , \alpha \rangle
 = R_X \, \langle Y , \V_X \alpha \rangle
\ee
and
\be
   \tilde{\V}_X (f \cdot Y) = (R_X f) \cdot \tilde{\V}_X Y
\ee
for a basic vector field $X$, by use of (\ref{XXdiscr-inv})
and (\ref{V_X-E}). In subsection~\ref{subsec:lcvf} we will see
that the inverse of $\tilde{\V}_X$, provided it exists, is
the parallel transport of a linear connection on $\X$,
associated with the linear connection on $\O^1$ in
a natural geometric way.

\vskip.2cm
\noindent
{\em Remark.} For a symmetric group lattice we may introduce
$\hat{\V}_{\ell_h} := \tilde{\V}_{\ell_{h^{-1}}}$ which satisfies
(\ref{Vellh_fE}) and thus defines a connection on $\X$
according to lemma~\ref{lemma:connE}.
\hfill $\blacksquare$

\subsection{The geometric meaning of (vanishing) torsion}
\label{subsec:Torsion=0}
For a biangle $h_1 h_2 = e$ the vanishing of the corresponding
part of the torsion 2-form (\ref{Thetah}) means
\be
     V^h{}_{h_1,h_2} = - \delta^h_{h_1}
\ee
and thus
\be
     \tilde{\V}_{\ell_{h_1}}(\ell_{h_2}) = - \ell_{h_1} \; .
\ee
We conclude that the transport $\tilde{\V}$ preserves a biangle
if the corresponding biangle torsion vanishes.
\vskip.2cm

For a triangle $h_1 h_2 = h_0$, the corresponding part of the
torsion 2-form (\ref{Thetah}) vanishes if and only if
\be
   V^h{}_{h_1,h_2} = \delta^h_{h_0} - \delta^h_{h_1}
\ee
which can be written as
\be
    \tilde{\V}_{\ell_{h_1}} ( \ell_{h_2} )
  = \ell_{h_0} - \ell_{h_1}  \; .  \label{triangleTors=0}
\ee
Associated with the latter triangle, there is a triangle
composed of the two vectors $\ell_{h_1}$ and $\ell_{h_0}$
at $g \in G$, and the vector $\ell_{h_2}$ at $g h_1$. The latter vector is
backwards parallel transported by $\tilde{\V}_{\ell_{h_1}}$
to the tangent space at $g$. The condition of vanishing
triangle torsion means that the three vectors at $g$ then
form a triangle. In this sense the transport preserves triangles
if the triangle torsion vanishes.
\vskip.2cm

A corresponding statement also holds for a quadrangle
$h_1 h_2 = \hat{h}_1 \hat{h}_2 = g \not\in S_e$. If we
consider\cite{2formcomp}
\be
   Q^h_{(g)\, h_1, h_2; \hat{h}_1, \hat{h}_2}
 = \check{Q}^h_{(g)\, h_1, h_2} - \check{Q}^h_{(g)\, \hat{h}_1, \hat{h}_2}
\ee
as the associated quadrangle torsion part, its vanishing means
\be
     V^h{}_{h_1,h_2} - V^h{}_{\hat{h}_1,\hat{h}_2}
   = \dl^h_{\hat{h}_1} - \dl^h_{h_1}   \label{quadrTors=0}
\ee
which is equivalent to
\be
    \tilde{\V}_{\ell_{h_1}} ( \ell_{h_2} )
    - \tilde{\V}_{\ell_{\hat{h}_1}} ( \ell_{\hat{h}_2} )
  = \ell_{\hat{h}_1} - \ell_{h_1}  \; .
\ee
There is a quadrangle composed of the two vectors $\ell_{h_1}$
and $\ell_{\hat{h}_1}$ at $g \in G$, the vector $\ell_{h_2}$ at
$g h_1$ and the vector $\ell_{\hat{h}_2}$ at $g \hat{h}_1$.
The latter two vectors are backwards parallel transported by
$\tilde{\V}_{\ell_{h_1}}$ and $\tilde{\V}_{\ell_{\hat{h}_1}}$,
respectively, to the tangent space at $g$. The condition of
vanishing quadrangle torsion then has the effect that the
resulting four vectors also form a quadrangle in the tangent
space at $g$.
\vskip.2cm

The presence of biangle, triangle, and quadrangle torsion thus
means that a biangle, triangle, and quadrangle composed of
vectors $\ell_h$ in the group lattice is, in general, not
mapped by $\tilde{\V}$ to a closed polygon in a tangent space.

\subsection{Linear connections on vector fields}
\label{subsec:lcvf}
Again, let $\na : \O^1 \rightarrow \O^1 \oA \O^1$ be a
linear connection with parallel transport operator $\V_X$.
Then
\be
   \langle \U_X Y , \V_X \alpha \rangle
 = R_X^{-1} \, \langle Y , \alpha \rangle
       \label{U_X-def}
\ee
for all basic vector fields $X$ associates with $\V_X$
a linear operator $\U_X : \X \rightarrow \X$. This definition
means that parallel transport preserves contractions of
vector fields and 1-forms. Writing
\be
    \U_{\ell_h} \ell_{h'}
  = \sum_{h'' \in S}
    R^\ast_{h^{-1}} (U_h)^{h''}{}_{h'} \cdot \ell_{h''}
       \label{Uell-ell}
\ee
with matrices $U_h$, we find from (\ref{U_X-def}) and
(\ref{Vellh_thetah}) that
\be
    U_h = V_h^{-1}  \; .
\ee
In particular, we need the $V_h$ to be invertible.
Furthermore, from (\ref{V_X-E}) we obtain
\be
  \U_X = \sum_{h \in S} (R_{h^{-1}}^\ast X^h) \cdot \U_{\ell_h} \; .
\ee

\begin{lemma}
\label{lemma:Ubas}
For basic vector fields $X$,
\be
   \U_X (f \cdot Y) = (R_X^{-1} f) \cdot \U_X(Y) \; .
       \label{U_basicX(fY)}
\ee
\end{lemma}
{\bf Proof:} Using (\ref{XXdiscr-inv}) and (\ref{R_Xbas_inv})
we obtain
\bez
     \langle \U_X (f \cdot Y) , \V_X \alpha \rangle
 &=& R_X^{-1} \, \langle f \cdot Y , \alpha \rangle
  = (R_X^{-1} f) \, R_X^{-1} \, \langle Y , \alpha \rangle
  = (R_X^{-1} f) \, \langle \U_X Y , \V_X \alpha \rangle  \\
 &=& \langle (R_X^{-1} f) \cdot \U_X Y , \V_X \alpha \rangle \; .
\eez
\hfill $\blacksquare$
\bigskip

In particular,
\be
  \U_{\ell_h}(f \cdot Y) = (R^\ast_{h^{-1}} f) \cdot \U_{\ell_h} Y
       \label{Uell(fY)}
\ee
so that
\be
   \na Y := \theta \oA Y - \U(Y) \, , \qquad
   \U(Y) := \sum_{h \in S} \theta^h \oA \U_{\ell_h} Y
\ee
defines a connection on $\X$, i.e. a linear map
$\na : \X \to \O^1 \oA \X$ with the property
$\na (f \cdot Y) = \d f \oA Y + f \, \na Y$ (see
section~\ref{sec:connections}). Next we establish the relation
with the transport $\tilde{\V}_X$ introduced in the previous
subsection.

\begin{lemma}
\label{lemma:UV}
For basic vector fields $X$,
\be
   \U_X = \tilde{\V}_X^{-1} \; .
\ee
\end{lemma}
{\bf Proof:} With the help of (\ref{tV(fY)}), (\ref{tVell-ell}),
(\ref{Uell-ell}) and (\ref{Uell(fY)}) we find
$\U_{\ell_h} \tilde{\V}_{\ell_h} Y = Y$ for all vector fields $Y$.
Using (\ref{XXdiscr-inv}) for an arbitrary basic vector field $X$
and $\sum_{h \in S} X^h = \mathbf{1}$, we obtain
\bez
      \U_X \tilde{\V}_X
  &=& \sum_{h,h' \in S} (R_{h^{-1}}^\ast X^h) \cdot
      \U_{\ell_h} ( X^{h'} \, \tilde{\V}_{\ell_{h'}} ) \\
  &=& \sum_{h,h' \in S} (R_{h^{-1}}^\ast X^h) \, (R_{h^{-1}}^\ast X^{h'}) \cdot
      \U_{\ell_h} \, \tilde{\V}_{\ell_{h'}}  \\
  &=& \sum_{h \in S} (R_{h^{-1}}^\ast X^h) \cdot \U_{\ell_h} \tilde{\V}_{\ell_h}
   = \sum_{h \in S} X^h \, I
   = I   \; .
\eez
\hfill $\blacksquare$
\bigskip

Let us also define
\be
   \na_{\ell_h} Y := Y - \U_{\ell_h} Y \, , \qquad
   \na_{\ell_h} \alpha := \alpha - \V_{\ell_h} \alpha
\ee
which satisfy
\be
  \na_{\ell_h} (f \, \alpha) = (\bar{\ell}_h f) \, \alpha
 + (R^\ast_{h^{-1}} f) \, \na_{\ell_h} \alpha \, , \quad
  \na_{\ell_h} (f \cdot Y) = (\bar{\ell}_h f) \cdot Y
 + (R^\ast_{h^{-1}} f) \cdot \na_{\ell_h} Y
\ee
where
\be
  \bar{\ell}_h = \ell_h \, R^\ast_{h^{-1}} = I - R^\ast_{h^{-1}}
      \label{barell_h}
\ee
is the backward difference operator on $\A$. Then the following identity
holds:
\be
   \bar{\ell}_h \langle Y , \alpha \rangle
 = \langle \na_{\ell_h} Y , \alpha \rangle
   + \langle Y , \na_{\ell_h} \alpha \rangle
   - \langle \na_{\ell_h} Y , \na_{\ell_h} \alpha \rangle \; .
\ee
\vskip.2cm

In general, the parallel transport of a discrete vector
field along a discrete vector field is not a discrete vector
field. A parallel transport or connection which maps
discrete vector fields into discrete vector fields will
be called ``discrete''. In this case, the matrices $V_h$
represent permutations.

\section{Differential calculi on coset spaces of discrete groups}
\label{sec:dc_coset}
\setcounter{equation}{0}
Let $H$ be a subgroup of $G$ and let $G/H$ denote the set of
right cosets of $H$ in $G$, i.e. $K \in G/H$ has the form $K=Hg$
for some $g \in G$. The algebra $\A_{G/H}$ of complex valued
functions $F : G/H \rightarrow \mathbb{C}$ can be naturally
identified with a subalgebra of the algebra $\A = \A_G$ of
functions on $G$.
Since the cosets form a partition of $G$, using our notation
(\ref{e^K}) we find $e^K e^{K'} = \dl^{K,K'} \, e^K$ and
$\sum_{K \in G/H} e^K = \mathbf{1}$. As a consequence, each
element $F \in \A_{G/H}$ has a unique decomposition
$F = \sum_{K \in G/H} F(K) \, e^K$. The right action of $G$
on $G$ induces a right action on $G/H$:
\be
   R_g^\ast F
 = \sum_{K \in G/H} F(K) \, R_g^\ast e^K
 = \sum_{K \in G/H} F(K) \, e^{K g^{-1}} \; .
\ee
\vskip.2cm

Let $(G,S)$ be a group lattice. The 1-forms $\{ \theta^h | \, h \in S\}$
then generate an $\A_{G/H}$-bimodule $\O^1_{G/H}$ such that
\be
   \theta^h \, e^K = (R^\ast_h \, e^K) \, \theta^h
                   = e^{K h^{-1}} \theta^h
\ee
and
\be
  \d e^K = \sum_{h \in S} (e^{Kh^{-1}} - e^K) \, \theta^h
\ee
defines a linear map $\d : \A_{G/H} \rightarrow \O^1_{G/H}$ which satisfies
\be
    \d (F F') = (\d F) \, F' + F \, \d F'
\ee
for all $F,F' \in \A_{G/H}$, so that $(\O^1_{G/H}, \d)$ is
a first order differential calculus over $\A_{G/H}$. It is important
to note, however, that $\O^1_{G/H}$ is not, in general, generated by
$\A_{G/H}$, i.e. $\A_{G/H} \, (\d \A_{G/H}) \, \A_{G/H}$ is smaller
than $\O^1_{G/H}$.\cite{gdc}
In any case, the first order differential calculus extends to
a differential calculus $(\O_{G/H},\d)$ over $\A_{G/H}$.
A closer inspection shows that the latter is simply obtained from the
group lattice differential calculus $(\O,\d)$ by restricting $\A_G$
to the subalgebra of functions corresponding to $\A_{G/H}$.
\vskip.2cm

Drawing an arrow from a point representing a coset $K$ to a point
representing a coset $K'$ whenever there is an $h \in S$ such that
$K' = Kh$, we obtain a digraph. This \emph{coset digraph}\cite{Ccd}
is also known as the {\em Schreier diagram} of the triple $(G,S,H)$
(see Ref. \cite{Bollo98}, for example).
In contrast to the digraphs (group lattices) considered in the
previous sections, coset digraphs may have multiple arrows between
two sites and even loops (i.e. arrows from a site to itself).
Indeed, whenever we have different
$h, h' \in S$ such that $H h = H h'$, then there are multiple
arrows in a coset digraph. The resulting discrete geometry is
therefore more complex than the one determined by (ordinary)
differential calculi on the algebra of functions on the
corresponding set of points. Such a generalization may prove to
be relevant for the description of electric circuits, for example.
\vskip.2cm

\begin{example}
\label{ex:Z2-coset}
Let $G = \mathbb{Z}_2 = \{ 0,1 \}$, the cyclic group of order 2 with
group operation the addition modulo 2. With $S = \{ 1 \}$, we have
$S_{(0)} = S$, $S_{(1)} = S_{(2)} = \emptyset$.
Choosing $H = G$, the coset space consists of a single element only
and the algebra of functions on it is therefore $\mathbb{C}$.
The 1-form $\theta^1$ corresponds to a loop (see Fig.~\ref{fig:z2}).
Then we have $\O^r_{\mathbb{Z}_2/\mathbb{Z}_2}
 = \mbox{span}_{\mathbb{C}}\{ (\theta^1)^r \}$ for $r > 0$.
According to (\ref{dthetah}), the action of $\d$ on forms is determined
by $\d \theta^1 = 2 \, \theta^1 \theta^1$ together with the Leibniz rule.
As a consequence,
$\d \theta^{2r} = 0$ and $\d \theta^{2r+1} = 2 \, (\theta^1)^{2(r+1)}$.
\begin{figure}
\begin{center}
\includegraphics[scale=.8]{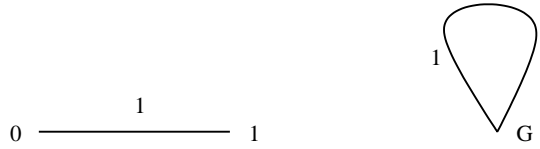}
\caption{The symmetric digraph of $(\mathbb{Z}_2,\{ 1 \})$ and
the corresponding coset graph corresponding to the choice
$H = \mathbb{Z}_2$.}\label{fig:z2}
\end{center}
\end{figure}
\end{example}

\begin{example}
\label{ex:Z3-coset}
Choosing $\mathbb{Z}_3$ with $S = \{ 1,2 \}$ and passing
to the single point coset space $\mathbb{Z}_3/\mathbb{Z}_3$,
one remains with two 1-forms $\theta^h$, $h=1,2$.
According to (\ref{dthetah}), the exterior derivative then
acts as follows,
\be
 \d \theta^1 &=& 2 \, (\theta^1)^2 - (\theta^2)^2
                 + \theta^1 \, \theta^2 + \theta^2 \, \theta^1  \\
 \d \theta^2 &=& 2 \, (\theta^2)^2 - (\theta^1)^2
                 + \theta^1 \, \theta^2 + \theta^2 \, \theta^1
                 \; .
\ee
\end{example}

\begin{example}
\label{ex:Z6-coset}
An example of a coset digraph containing both loops and multiple links
is given by $G = \mathbb{Z}_6$, the cyclic group of order 6 with group
operation $\dotplus$, addition modulo 6. With $S = \{ 1,2,3 \}$ we find
$S_{(0)} = \{ 3 \}$, $S_{(1)} = \{ 2,3 \}$ and $S_{(2)} = \{ 4,5 \}$
which implies that we have two 2-form relations and consequently seven
independent 2-forms.
Choosing $H = \{ 0,2,4 \}$, there are only two cosets, $H$ and $1 \dotplus H$.
Since $H \stackrel{1,3}{\longleftrightarrow} 1 \dotplus H$,
$H \stackrel{2}{\longleftrightarrow} H$ and $1 \dotplus  H \stackrel{2}{\longleftrightarrow}
1 \dotplus H$, we obtain the graph in Fig.~\ref{fig:z6}.
\begin{figure}
\begin{center}
\includegraphics[scale=.8]{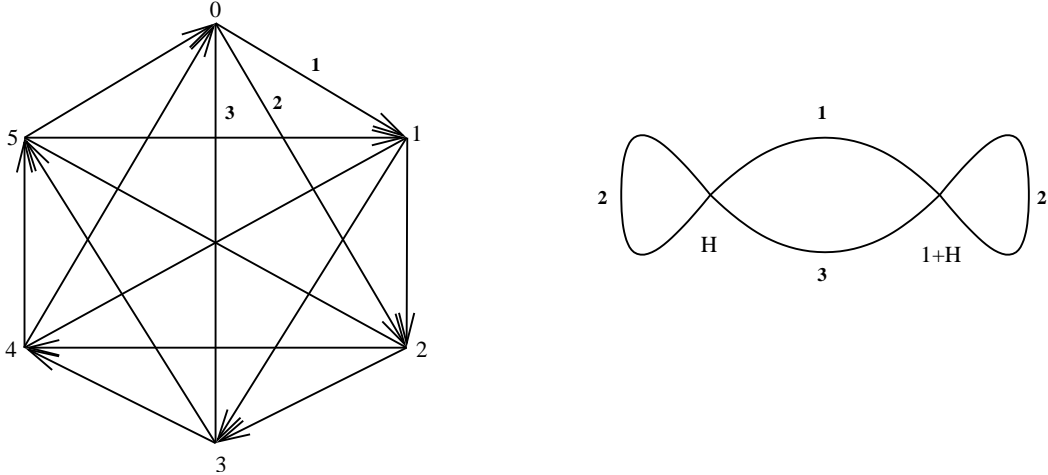}
\caption{Group lattice of $(\mathbb{Z}_6, \{ 1,2,3 \})$ and the
coset digraph for $H = \{ 0,2,4 \}$.}\label{fig:z6}
\end{center}
\end{figure}
\end{example}

\begin{example}
\label{ex:S3-coset}
Let us consider the ${\cal S}_3$ group lattice
of examples~\ref{ex:G=S3} and \ref{ex:S3-2forms}
with $S = \{ (12),(13),(23) \}$.
The following table expresses the action of $R_h$, $h \in S$,
on $G$:
\bez
 \begin{array}{r|ccc}
  g \backslash h & (12)  & (13)  & (23)  \\ \hline
               e & (12)  & (13)  & (23)  \\
            (12) &  e    & (123) & (132) \\
            (13) & (132) &   e   & (123) \\
            (23) & (123) & (132) &   e   \\
           (123) & (23)  & (12)  & (13)  \\
           (132) & (13)  & (23)  & (12)
 \end{array}
\eez
Choosing the subgroup $H = \{ e,(12) \}$, the corresponding right
cosets are $H$, $H(13) = \{ (13),(123) \}$ and $H(23) = \{ (23),(132) \}$
(see Fig.~\ref{fig:s3cosets}).
\begin{figure}
\begin{center}
\includegraphics[scale=.8]{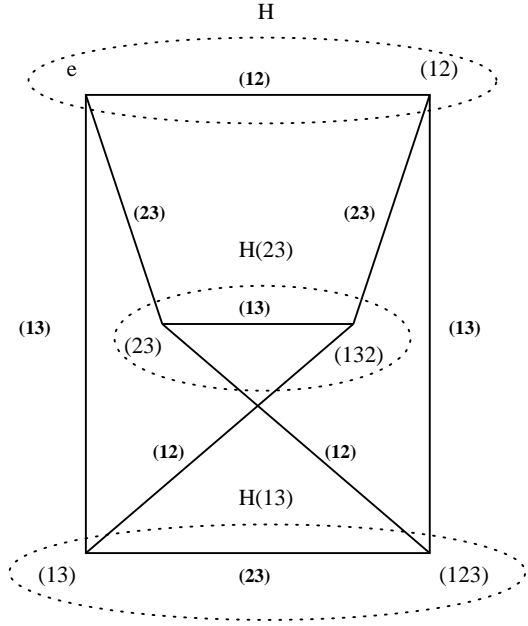}
\caption{The right cosets of ${\cal S}_3$ with respect to the
subgroup $H = \{ e,(12) \}$.}\label{fig:s3cosets}
\end{center}
\end{figure}

The action of $R_h$, $h \in S$, on the cosets is given by the
following table:
\bez
 \begin{array}{r|ccc}
   H g \backslash h & (12)  & (13)  & (23)  \\ \hline
                 H &  H    & H(13) & H(23) \\
             H(13) & H(23) & H     & H(13) \\
             H(23) & H(13) & H(23) & H
 \end{array}
\eez

Since $H \stackrel{(12)}{\mapsto} H$,
$H(13) \stackrel{(23)}{\mapsto} H(13)$
and $H(23) \stackrel{(13)}{\mapsto} H(23)$,
there are loops in the coset digraph (see Fig.~\ref{fig:s3H-loops}).
This has its origin in the fact that $S \cap H = \{ (12) \}$
and thus $H(12) = H$.\cite{loops}
The 1-forms $e^H \theta^{(12)}$, $e^{H(13)} \theta^{(23)}$ and
$e^{H(23)} \theta^{(13)}$ are associated with the loops and therefore
cannot be expressed in terms of functions and differentials.
In order to eliminate the loops, one could set these 1-forms to zero.
As a consequence of such additional relations, the resulting bimodule
of 1-forms is no longer free.

\begin{figure}
\begin{center}
\includegraphics[scale=.7]{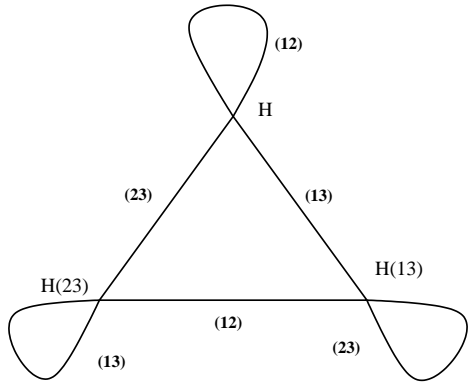}
\caption{Loops in the coset digraph of ${\cal S}_3$ with respect
to the subgroup $H = \{ e,(12) \}$ and
$S = \{ (12),(13),(23) \}$.}\label{fig:s3H-loops}
\end{center}
\end{figure}

\begin{figure}
\begin{center}
\includegraphics[scale=.8]{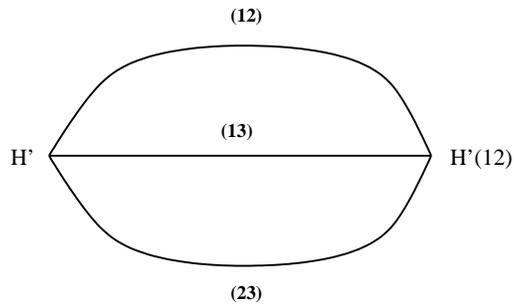}
\caption{An ${\cal S}_3$ coset digraph with multiple
arrows.}\label{fig:s3HH}
\end{center}
\end{figure}

As a further example, let us consider the subgroup
$H' = \{ e,(123),(132) \}$. The corresponding cosets are $H'$
and $H'(12) = \{ (12),(23),(13) \}$.
The table of the right action $R_h$ on these cosets is then
\bez
 \begin{array}{r|ccc}
  H'g \backslash h & (12)   & (13)   & (23)   \\ \hline
                H' & H'(12) & H'(12) & H'(12) \\
            H'(12) &  H'    & H'     & H'
 \end{array}
\eez
In this case, we have multiple arrows in the coset digraph
(see Fig.~\ref{fig:s3HH}). By imposing the relations
$\theta^{(12)} = \theta^{(13)} = \theta^{(23)}$ on the
differential calculus, we could eliminate the multiple links.
\end{example}

\begin{example}
\label{ex:S4-coset}
Let $G = {\cal S}_4$ and
$S = \{ (12),(13),(14),(23),(24),(34) \}$, as in
example~\ref{ex:S4-2forms}. Furthermore, we choose a subgroup $H$
of order 3 with eight cosets:
\bez
\begin{array}{rclcrcl}
         H &=& \{ e,(123),(132) \}  &\qquad& H(12) &=& \{ (12),(23),(13) \}     \\
 H(12)(34) &=& \{ (12)(34),(243),(143) \} && H(14) &=& \{ (14),(1234),(1324) \} \\
 H(13)(24) &=& \{ (142),(234),(13)(24) \} && H(24) &=& \{ ((24),(1423),(1342) \}\\
 H(14)(23) &=& \{ (124),(14)(23),(134) \} && H(34) &=& \{ (34),(1243),(1432) \}
 \, .
\end{array}
\eez
The table of right actions of the elements of $S$ on $G/H$ is
\bez
\begin{array}{r|cccccc} Hg\backslash h&(12)&(13)&(14)&(23)&(24)&(34)\\ \hline
 H         &H(12)     &H(12)     &H(14)     &H(12)     &H(24)     &H(34)     \\
 H(12)     &H         &H         &H(14)(23) &H         &H(13)(24) &H(12)(34) \\
 H(14)     &H(13)(24) &H(12)(34) &H         &H(14)(23) &H(14)(23) &H(14)(23) \\
 H(24)     &H(14)(23) &H(13)(24) &H(13)(24) &H(12)(34) &H         &H(13)(24) \\
 H(34)     &H(12)(34) &H(14)(23) &H(12)(34) &H(13)(24) &H(12)(34) &H         \\
 H(12)(34) &H(34)     &H(14)     &H(34)     &H(24)     &H(34)     &H(12)     \\
 H(13)(24) &H(14)     &H(24)     &H(24)     &H(34)     &H(12)     &H(24)     \\
 H(14)(23) &H(24)     &H(34)     &H(12)     &H(14)     &H(14)     &H(14)
\end{array}
\eez
and the coset digraph is drawn in Fig.~\ref{fig:s4H}.
\begin{figure}
\begin{center}
\includegraphics[scale=.6]{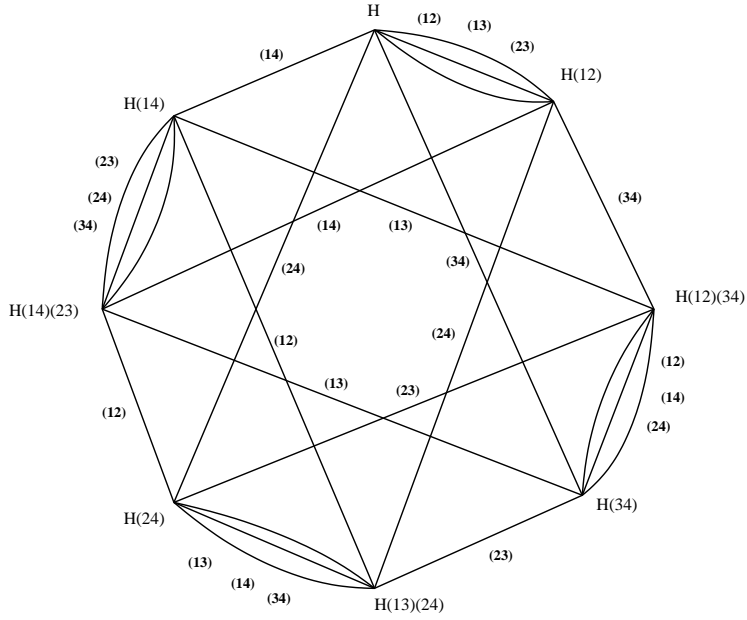}
\caption{Coset digraph of ${\cal S}_4$ with $S = \{ (12),(13),(14),(23),(24),(34) \}$
and $H = \{ e,(123),(132) \}$.}\label{fig:s4H}
\end{center}
\end{figure}
If we impose the relations
\bez
 e^H \theta^{(12)} = e^H\theta^{(13)} = e^H\theta^{(23)} \, , \qquad
 e^{H(12)} \theta^{(12)} = e^{H(12)} \theta^{(13)}
                         = e^{H(12)} \theta^{(23)} \,, \\
 e^{(12)(34)H} \theta^{(12)} = e^{H(12)(34)} \theta^{(14)}
                             = e^{H(12)(34)} \theta^{(24)} \, , \quad
 e^{H(34)} \theta^{(12)} = e^{H(34)} \theta^{(14)}
                         = e^{H(34)} \theta^{(24)} \,, \\
 e^{(13)(24)H} \theta^{(13)} = e^{H(13)(24)} \theta^{(14)}
                             = e^{H(13)(24)} \theta^{(34)} \, , \quad
 e^{H(24)} \theta^{(13)} = e^{H(24)} \theta^{(24)}
                         = e^{H(24)} \theta^{(34)} \, , \\
 e^{(14)(23)H} \theta^{(23)} = e^{H(14)(23)} \theta^{(24)}
                             = e^{H(14)(23)} \theta^{(34)} \, , \quad
 e^{H(14)} \theta^{(23)} = e^{H(14)} \theta^{(24)}
                         = e^{H(14)} \theta^{(34)}
\eez
then the multiple links are eliminated. The bimodules of differential forms
are then no longer free.
\end{example}

The relations one has to impose on the 1-forms of a generalized differential
calculus on a coset space in order to reduce it to an ordinary differential
calculus (without loops or multiple links in the associated digraph) are
of the form $e^K \theta^h = 0$ or $e^K (\theta^{h_1} - \theta^{h_2}) = 0$.
Such relations do not lead to additional higher form relations.
For relations eliminating loops this follows from
\bez
     \d(e^K \theta^h) = (\theta e^K - e^K \theta) \, \theta^h
     + e^K \, (\theta \, \theta^h + \theta^h \, \theta - \Delta(\theta^h))
   = - e^K \Delta(\theta^h)
   = - \Delta(e^K \theta^h) \, .
\eez
A similar calculation applies to relations eliminating multiple links.

\subsection{Higgs field from gauge theory with an internal coset lattice}
\label{subsec:coset-Higgs}
Let $(\O,\d)$ be the usual differential calculus over the algebra $\A$ of smooth
functions on $\mathbb{R}^n$. Furthermore, let $(\tilde{\O},\tilde{\d})$ denote
the ``loop'' differential calculus over the algebra $\tilde{\A} = \mathbb{C}$ of
functions on the single point space $\mathbb{Z}_2/\mathbb{Z}_2$, see
example~\ref{ex:Z2-coset}. The skew-tensor product\cite{skew}
$\hat{\O} = \O \hat{\otimes} \tilde{\O}$ of the two differential
calculi then defines a new differential calculus $(\hat{\O},\hat{\d})$
over $\A$.
\vskip.1cm

Let us introduce $\rho :=(1/c) \, \theta^1$ with a real constant $c$,
so that $\tilde{\d} \rho = 2 c \, \rho^2$, $\tilde{\d} \rho^{2r} = 0$ and
$\tilde{\d} \rho^{2r+1} = 2 c \, \rho^{2r+2}$ (see example~\ref{ex:Z2-coset}).
Then $\hat{\d} f = \d f$ and $\hat{\d}(\omega \rho^r) =
(\d \omega) \rho^r + (-1)^s \, \omega \, \tilde{\d} \rho^r$ for
$\omega \in \O^s$.
\begin{figure}
\begin{center}
\includegraphics[scale=.7]{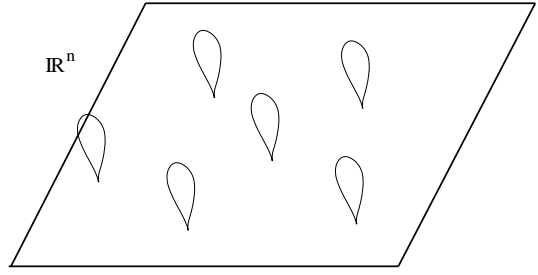}
\caption{Visualization of the geometry considered in
section~\ref{subsec:coset-Higgs}.}\label{fig:carpet}
\end{center}
\end{figure}
Let $\hat{A}$ be a gauge potential 1-form.
With the decomposition $\hat{A} = A + \phi \, \rho$,
the field strength $\hat{F} = \hat{\d} \hat{A} + \hat{A}^2$
becomes
\be
   \hat{F} = F + \D \phi \, \rho + (\phi^2 + 2c \, \phi) \, \rho^2
\ee
where we used $A \rho = - \rho A$ and introduced the exterior covariant
derivative $\D \phi = \d \phi + [A,\phi]$.
In terms of $\varphi := \phi + c \, I$, this reads
\be
   \hat{F} = F + \D \varphi \, \rho + (\varphi^2 - c^2 I) \, \rho^2 \, .
\ee
Let us now introduce an inner product on $\hat{\O}$ such that
\be
    (\omega \rho^r , \omega' \rho^s)
 := \dl_{rs} \, \lambda^r \, (\omega , \omega')
\ee
with a positive constant $\lambda$ and the usual sesquilinear inner
product\cite{ip_Rm} $(\omega , \omega')$ of differential forms on
$\mathbb{C}^n$ with respect to a (pseudo-) Riemannian metric.
Then we find
\be
    (\hat{F} , \hat{F})
  = {1 \over 2} F^\dagger_{\mu\nu} F^{\mu\nu}
    + \lambda \, (\na_\mu \varphi)^\dagger \, \na^\mu \varphi
    + \lambda^2 \, (\varphi^2 - c^2 I)^\dagger \, (\varphi^2 - c^2 I) \, .
     \label{in_YM}
\ee
If we set
\be
  \varphi = \left(\begin{array}{cc}
                  0 &  \chi^\dagger  \\
               \chi & 0
                  \end{array}\right)
\ee
then
\be
  \varphi^2 =  \left(\begin{array}{cc}
                 \chi^\dagger \chi & 0 \\
                                 0 & \chi \chi^\dagger
                     \end{array} \right) \, .
\ee
Taking the trace of (\ref{in_YM}) results in
\be
    {\rm tr}(\hat{F} , \hat{F})
  = {1 \over 2} \, {\rm tr} (F^\dagger_{\mu\nu} F^{\mu\nu})
    + 2 \, \lambda \, {\rm tr} \Big((\na_\mu \chi)^\dagger \na^\mu\chi\Big)
    + 2 \, \lambda^2 (\|\chi\|^2 - c^2 I)^2 \, .
\ee
The constants can now be chosen in such a way that the usual
Yang-Mills-Higgs Lagrangian is obtained.
More complicated examples can be constructed by replacing
$\mathbb{Z}_2$ with $\mathbb{Z}_N$, $N>2$ (see also
example~\ref{ex:Z3-coset} and Ref. \cite{Okum96}).

\section{Conclusions}
\label{sec:concl}
\setcounter{equation}{0}
With this work we have started to develop a formalism of
differential geometry of group lattices, based on elementary
concepts of non-commutative geometry.
A group lattice $(G,S)$ naturally determines a differential calculus
over the algebra of functions on the discrete group $G$ and we
systematically explored the structure of differential calculi
which emerge in this way.
\vskip.2cm

Counterparts of the Yang-Mills action on arbitrary group lattices
have been obtained. They generalize the familiar action of lattice
gauge theory. In particular, these can be further analyzed using
the methods of Ref. \cite{Lech+Samu95}.
\vskip.2cm

Whereas noncommutative geometry conveniently defines general geometric
structures in terms of differential forms, their geometric
significance in special cases, like the group lattices under
consideration, is often easier to understand when expressed in
terms of vector fields.
A large part of this work has therefore been devoted to the
properties of a class of vector fields on group lattices, which
we called ``discrete vector fields'', and the subclass of
``basic vector fields''. We also introduced an inner product of
discrete vector fields (with differentiable flow) and forms.
In particular, this opens the possibility to develop mechanics
on group lattices using familiar formulae of symplectic geometry.
\vskip.2cm

A linear connection (on the space of 1-forms) on a group lattice
defines a parallel transport of vector fields along a vector field.
We found a very simple geometric picture associated with the
condition of vanishing torsion, which strongly corroborates
the formalism.
\vskip.2cm

Continuing this work, in a forthcoming paper we develop ``Riemannian geometry''
on group lattices. More precisely, for making contact with classical geometry,
the subclass of {\em bicovariant} group lattices turns out to be distinguished.
We introduced these lattices as those for which all the left and right actions
$L_h, R_h$, $h \in S$, are differentiable maps (in the sense of
section~\ref{sec:diffmaps}).
\vskip.2cm

The geometric framework presented in this work may also be helpful for
the construction and analysis of completely integrable models on
group lattices. The differential calculus associated with a linear
or quadratic lattice (see example~\ref{ex:G=Z,S=1}) has already
been applied in this context \cite{DMH96}.

\begin{appendix}
\renewcommand{\theequation} {\Alph{section}.\arabic{equation}}
\addcontentsline{toc}{section}{\numberline{}Appendices}

\section{Integral curves of discrete vector fields}
\label{sec:int_curv}
\setcounter{equation}{0}
Let $(G,S)$ be a group lattice.
A map $\gamma : \mathbb{Z} \to G$ which is a solution of the equation
\be
  \pa_{+t}(\gamma^\ast f) = \gamma^\ast(X f)
        \qquad ( \forall f \in \A )
\ee
for some discrete vector field $X = \sum_{h \in S} X^h \cdot \ell_h \in \X$
is called an {\em integral curve} of $X$. More explicitly, this reads
\be
   f(\gamma(t+1)) - f(\gamma(t))
 = \sum_{h \in S} X^h(\gamma(t)) \, [ f(\gamma(t)h) - f(\gamma(t)) ]
\ee
or
\be
  f(\gamma(t+1)) &=& \sum_{h \in S} X^h(\gamma(t)) \, f(\gamma(t)h)
  + \Big( 1 - \sum_{h \in S} X^h(\gamma(t)) \Big) f(\gamma(t))
                \nonumber \\
  &=& \sum_{h \in S_e} X^h(\gamma(t)) \, f(\gamma(t) h)
   = \Big( (I+X)f \Big)(\gamma(t))  \label{f_dint_curve}
\ee
(where $X^e(g)=1$ iff $X^h(g)=0$ for all $h \in S$).
Since precisely one component $X^h(\gamma(t))$, $h \in S_e$, is
different from zero and then equal to 1, we obtain
$ f(\gamma(t+1)) = f( \sum_{h \in S_e} X^h(\gamma(t)) \, \gamma(t) h)$
for all $f \in \A$ and thus
\be
  \gamma(t+1) = \sum_{h \in S_e} X^h(\gamma(t)) \, \gamma(t) h \; .
     \label{dint_curve}
\ee
The flow $\phi_t : G \to G$ generated by $X$ has to satisfy the
same equation, so that
\be
  \phi_{t+1} = \sum_{h \in S_e} (X^h \, R_h) \circ \phi_t \, .
\ee
Furthermore, $\phi_0 = \mbox{id}$, the identity on $G$.
On functions, we have (cf (\ref{f_dint_curve}))
\be
    \phi^\ast_{t+1} f = \phi_t^\ast ((I+X) f)
\ee
with the solution
\be
   \phi^\ast_t = (I+X)^t
\ee
as expected on the basis of our earlier considerations.
\vskip.1cm

Let us supply $\mathbb{Z}$ with the first order differential calculus
of example~\ref{ex:G=Z,S=1}, and $G$ with the calculus
associated with the subset $S \subset G \setminus \{ e \}$. According
to the criterium (\ref{phi_diff}), the map $\gamma$ is differentiable
iff $\gamma(t)^{-1} \gamma(t+1) \in S_e$ for all $t \in \mathbb{Z}$.
But this is automatically satisfied for an integral curve as
a consequence of (\ref{dint_curve}). We have already learned, however,
that the flow of $X$ is not in general differentiable as a map
$G \rightarrow G$ (with respect to the differential calculus
induced by $S$).

\end{appendix}


\begin{thebibliography}{99}
\bibitem{Cayley} We do not distinguish between the Cayley graph
 and its representation by points and arrows, the Cayley diagram.
\bibitem{genform} The formalism developed in this work may be
 generalized by replacing $\mathbb{C}$ with an arbitrary field $\mathbb{K}$.
\bibitem{notation} Here and in the following we slightly depart from
 corresponding definitions used in Ref. \cite{BDMHS96}.
 In particular, our definition of $\theta^h$ below corresponds
 to $\theta^{h^{-1}}$ in the latter work.
\bibitem{subbimodule} This is a sub-bimodule of the bimodule generated
 by all $\ell_g$, $g \in G$. Of course, it depends on the choice of $S$
 and should therefore rather be denoted as $\X_S$, for example.
 Since in most parts of this paper $S$ is a fixed set, we
 simply write $\X$ instead of $\X_S$.
\bibitem{G-infinite} If $|G|$ is infinite, the right side
 looks like an infinite sum. Note, however, that at each element
 $g \in G$ this reduces to $\theta^h|_g = \d e^{gh}|_g$.
\bibitem{combin} A Cayley graph without biangles (circuits of length 2)
 is sometimes called ``combinatorial'', see Ref. \cite{Serr80}.
\bibitem{Delta-cons} The consistency condition
 $\Delta (\theta^h) \, f = \Delta (\theta^h \, f)
  = \Delta (\R_h^\ast f \, \theta^h) = \R_h^\ast f \, \Delta(\theta^h)$
 is indeed satisfied as a consequence of (\ref{Delta}) and (\ref{theta_f}).
\bibitem{d-notri} If there are no triangles, then we have simply
 $\d \alpha = \theta \alpha + \alpha \theta$ for a 1-form $\alpha$.
\bibitem{intertw} This means that $\phi_X^\ast$ intertwines
 the corresponding maps associated with discrete vector fields.
 Note that, in general, $Z$ is not unique.
\bibitem{Xarrow} These conditions mean that if there is an outgoing
 $X$-arrow at $g$, then there is also precisely one incoming $X$-arrow.
 If there is no outgoing arrow, then there is also no incoming arrow.
\bibitem{bvS3} In fact, all basic vector fields on ${\cal S}_3$
 with $S = \{ (12),(13),(23) \}$ generate differentiable maps
 $G \rightarrow G$. This follows from $S^{-1} = S$ and $S^3 = S$,
 which implies that every expression of the form $s(g)^{-1} h s(gh)$
 (for $s$ with values in $S$) lies in $S$, so that condition (2) of
 theorem~\ref{theor:phiXdiff} holds.
\bibitem{innprod} Except for 1-forms, we have in general
 $X \lc \omega \neq \sum_h X^h \, (\ell_h \lc \omega)$ where
 $X = \sum_h X^h \cdot \ell_h$.
\bibitem{loops} Note that $Hh = H$ implies $(Hg) \ad(g^{-1})h = Hg$,
 so we also have a loop at $Hg$ and then at every coset. If $\ad(g)h \in H$
 and thus $Hgh = Hg$ for all $g$, we can eliminate these loops by reducing
 $S$ to $S \setminus \{ h \}$. But if $\ad(g)h \not\in H$ for some $g$,
 it will not be possible to get rid of the loops by choosing a smaller
 set $S$ without simultaneously eliminating some arrows between different
 points.
\bibitem{ip_Rm} It has the properties
 $(f \omega, f' \omega') = f^\dagger f' \, (\omega,\omega')$ and
 $(\omega,\omega')^\dagger = (\omega',\omega)$
 where ${}^\dagger$ denotes complex or hermitian conjugation.
 Furthermore,
 $(\d x^{\mu_1} \cdots \d x^{\mu_r} , \d x^{\nu_1} \cdots \d x^{\nu_s})
  = \dl_{r,s} \, g^{\mu_1 \rho_1} \cdots g^{\mu_r \rho_r} \,
    \dl^{[\nu_1}_{\rho_1} \cdots \dl^{\nu_r]}_{\rho_r}$.
\end{thebibliography}
\end{document}